\documentclass[12pt]{iopart}

\usepackage{graphicx}
\usepackage{dcolumn}
\usepackage{bm}
\usepackage{subfigure}
\usepackage{amssymb}
\usepackage{float}
\usepackage{multirow}

\newcolumntype{.}{D{.}{.}{4}}
\newcolumntype{p}{D{(}{(}{2}}

\newcommand{\Mpl}{M_{\rm Pl}}

 \usepackage{epstopdf}

\newcommand{\iDens}{\,{\rm h^3\,Mpc^{-3}}}
\newcommand{\iMpc}{\,h{\rm\,Mpc^{-1}}}
\newcommand{\ud}{{\rm d}}
\newcommand{\tw}{\tilde w}

\renewcommand{\thesubfigure}{(\roman{subfigure})}
\makeatletter
\renewcommand{\@thesubfigure}{\thesubfigure\space}
\renewcommand{\p@subfigure}{\thefigure}
\makeatother
\def\bdm{\begin{displaymath}}
\def\edm{\end{displaymath}}

\begin{document}

\title{Inflation and the Scale Dependent Spectral Index: Prospects and Strategies }
\author{Peter Adshead, Richard Easther,}
\address{Department of Physics, Yale University, New Haven,  CT 06520, USA}

\author{Jonathan Pritchard\footnote{Hubble Fellow} and Abraham Loeb}
\address{Harvard-Smithsonian Center for Astrophysics, Cambridge, MA 02138, USA} 
\date{\today}

\begin{abstract} 
We consider the running of the spectral index  as a probe of both inflation itself, and of the  overall evolution of the very early universe.     Surveying a collection of simple single field inflationary models, we confirm that  the magnitude of the running is relatively consistent, unlike the tensor amplitude, which varies by orders of magnitude. Given this target,  we confirm that the running is potentially detectable by future large scale structure or 21 cm observations, but that  only the most futuristic measurements can distinguish between these models on the basis of their running.  For any specified inflationary scenario, the combination of the running index and unknown  post-inflationary expansion history induces a theoretical uncertainty in the predicted value of the spectral index. This effect can easily dominate the statistical uncertainty with which Planck and its successors are expected to measure the spectral index.  More positively, upcoming cosmological experiments thus provide an intriguing probe of physics between TeV and  GUT scales by constraining the reheating history  associated with any specified inflationary model, opening a window into the ``primordial dark age'' that follows the end of inflation.
\end{abstract}

\maketitle


\section{Introduction}

Concordance cosmology fits the overall properties of the universe in terms of a handful of physical parameters.   This set is not fixed, but is the minimal number required to match the data  \cite{Liddle:2004nh,Adshead:2008ky,Baumann:2008aq}. The concordance  parameter set will expand as more subtle physical effects are resolved in the data, while current variables  may someday be determined independently of cosmological observations. Within the concordance model we may delineate  {\em sectors\/}, or subsets of related parameters. For example, $\Omega_b$, $\Omega_{c}$ and $\Omega_\Lambda$  (the relative contributions of baryons, cold dark matter and dark energy to the present-day density of the universe) define the  composition of the universe. These quantities  stand in for our ignorance of aspects of fundamental physics: $\Omega_b$, $\Omega_{c}$ and $\Omega_\Lambda$ are free parameters because we do not understand  baryogenesis, the dark matter abundance, and the amplitude of the vacuum energy. 

Two concordance parameters specify the primordial perturbation spectrum -- the amplitude $A_s$ and spectral index $n_s$.       If  perturbations are generated during inflation, $A_s$ and $n_s$  form the nucleus of an {\it inflationary sector\/}.   Many further observables are related to the primordial perturbations, including tensor modes (their amplitude relative to the scalar perturbations $r$, and their spectral dependence, $n_t$), non-Gaussianity, running in the scalar spectral index ($\alpha_s = dn_s/d\log{k}$, where $k$ is comoving wavenumber), features in the power spectrum, and relics generated at the end of the inflationary epoch.   These ``optional'' observables are {\em fingerprints\/} of specific inflationary scenarios, in that {\em most\/} models predict that {\em most\/} of these parameters are very small:  finding a non-zero value for any one of these quantities would slash the number of viable inflationary models.
  
 Predictions for $n_s, \alpha_s, r$, {\em et al.\/} define a mapping from the fundamental physical description of inflation (namely, the gravitational action and relevant field content, which is the {\em inflationary sector\/} of very high energy  particle physics) into the parameter space of observables.  The canonical example of this mapping is   the ``zoo-plot'' which locates single field, slow roll inflationary models in the $(n_s,r)$ plane \cite{Dodelson:1997hr}.  Unfortunately, the clarity of this plot is partially due to its mapping a subset of   inflationary models  into a subset of inflationary observables.    Furthermore,  as we explore below,  our lack of knowledge of the post-inflationary expansion history renders this mapping inherently imprecise. 

The first WMAP data release marked a turning point in constraints on inflationary scenarios \cite{Peiris:2003ff}.  A host of currently ongoing and planned observations, looking at both the CMB and complementary datasets, ensures  that the {\em de facto\/} observational campaign to constrain the inflationary era will last for at least the next several decades.    The purpose of this paper is to highlight the importance of the running, or scale dependence of the spectral index, as a key inflationary parameter.   Our first task is to  explore the correlation between $\alpha_s$ and the measured value of  $n_s$ induced by the (unknown) post-inflationary expansion history: the equation of state for the post-inflationary universe determines how rapidly modes reenter the horizon, and thus the moment during inflation when a given mode  {\em leaves\/} the horizon.  By definition, $\alpha_s$  fixes the scale dependence of $n_s$, so the observed value of $n_s$ depends on  $\alpha_s$ via the post-inflationary expansion history.  Consequently,  assumptions about  the post-inflationary expansion history of the universe add a (sometimes implicit) theoretical prior to constraints on inflationary parameters. 

We consider several explicit inflationary models, namely natural inflation (a sinusoidal potential) \cite{Freese:1990rb}, $\phi^n$ inflation, where $n$ is a continuous parameter (motivated by models for which $n$ takes fractional values \cite{McAllister:2008hb,Flauger:2009ab}), and ``inverted'' potentials of the general form $\Lambda^4 - \lambda M_{\rm pl}^{4-n} \phi^n/n$ \cite{Kinney:1995cc,Kinney:1998md,Easther:2006qu}.   This is not an exhaustive survey, but it captures many well-known models.      We consider the theoretical uncertainty in predictions for inflationary observables via the {\em matching equation\/}, which connects comoving scales in the present universe to the moment during inflation at which the corresponding perturbations were produced. The matching depends on the rate at which modes reenter the horizon, which is in turn a function of the  post-inflationary expansion history. However, there are  few constraints on the detailed expansion history of the universe prior to nucleosynthesis, and the details of this epoch substantially modify the predictions of explicit inflationary models.

We consider three specific scenarios for the post-inflationary expansion.  The first is instant reheating, for which the inflationary observables are well defined. Secondly, we allow a matter dominated phase of unknown duration, followed by thermalization.  Finally, we consider a post-inflationary equation of state  fixed only by general physical arguments and direct observational constraints.   The unknown expansion history leads to an uncertainty,  $\Delta N$, in the number of e-folds,  $N$.  For  simple inflationary scenarios,   $10^{-4} \lesssim |\alpha_s|\lesssim 10^{-3}$, providing a benchmark value for this parameter. For a given  $\Delta N$, the measured value of $n_s$ changes by $\Delta n_s \sim\alpha_s \Delta N$.  Assuming a long period of matter domination before thermalization yields  $\Delta N \sim 10$ and $\Delta n_s \sim 0.005$ (similar to the statistical errors expected from Planck), and $\Delta N$ and $\Delta n_s$ can be much larger if we drop all theoretical priors on the post-inflationary dynamics.   Note that   this uncertainty is directly induced by the running: by definition, if $\alpha_s =0$, $n_s$ is not a function of $k$, and its value at the pivot scale is unaffected by the details of reheating.

Despite the large uncertainty in $n_s$ induced by the interaction between the running and post-inflationary expansion,  $\alpha_s$ is undetectable by   a cosmic variance limited CMB mission for simple inflationary models.   However,  the next decade will see dramatic improvement in the spatial volume and redshift range probed by large scale structure observations, and the beginning of the exploration of the dark ages via high redshift 21~cm emission. These measurements are directly sensitive to the primordial spectrum over a broad range of scales, and can improve limits on $\alpha_s$.  Unlike the CMB,  probes of large scale structure sample the universe volumetrically, and are not subject to Silk damping which washes out the CMB at small scales.   We explore the sensitivity of galaxy surveys and 21~cm experiments to $\alpha_s$, as  a function of their basic design parameters, confirming that ambitious surveys can detect  $\alpha_s$ at the level predicted by simple inflationary models. However, even highly ambitious proposals would have trouble measuring $\alpha_s$ with enough precision to distinguish otherwise degenerate inflationary scenarios.  On the other hand, the correlation between $n_s$ and the post-inflationary expansion history induced by the running means that  measurements of $n_s$ yield information about the post-inflationary expansion history following a {\em specified\/} inflationary scenario. This effectively uses the theoretical uncertainty in $n_s$ to gain information about the post-inflationary equation of state for a given inflationary model. While this is  a single number that depends on the integrated expansion history of the universe, it probes physical processes that occur between the TeV an GUT scales, which are otherwise almost unconstrained.     Specifically, this number yields information about the mechanism of post-inflationary reheating, and whether a specific inflationary model is compatible with -- or requires -- a transient matter dominated phase in the early universe, as well as more exotic scenarios such as kination \cite{Chung:2007vz} (or deflation \cite{Spokoiny:1993kt}) or a phase of cosmic string dominated evolution  \cite{Burgess:2005sb}.  
    
Given that we are considering simple, single field models,  our analysis touches on a great deal of earlier work. Large, negative values of $\alpha_s$  signal either the imminent breakdown of slowroll, or that higher order terms in the potential dominate the inflationary evolution \cite{Malquarti:2003ia,Easther:2006tv,Makarov:2005uh}.  Conversely, a positive running can be associated with primordial black hole production \cite{Kohri:2007qn,Peiris:2008be}, leading to further observational constraints. The correlation between the post-inflationary thermal history and inflationary observables is usually expressed via the matching equation, as surveyed in \cite{Liddle:2003as, Dodelson:2003vq,Alabidi:2005qi}.  Our analysis overlaps with discussions of the ``likely'' values of inflationary parameters: in particular, the natural value of $r$ in inflation is hotly debated \cite{Boyle:2005ug,Boyle:2008ri,Easther:2002rw,Bird:2008cp} while Kinney and Riotto have discussed the theoretical uncertainty in $n_s$ arising from  higher order terms  in the inflationary potential \cite{Kinney:2005in}.  Our treatment of the joint constraints on $\alpha_s$ from CMB, 21cm and surveys is complementary to treatments of combinations of CMB and gravitational wave experiments \cite{Kuroyanagi:2009br}.    Further the ability of future 21~cm experiments to detect a running index was analyzed by \cite{Barger:2008ii}, while \cite{Fedeli:2010iw} discusses the detectability of a running index through its effects on structure formation.
  
The paper is organized as follows. In section \ref{inflandpower} we review canonical single field inflation. We describe the matching of physical scales to field values during inflation and introduce the effective equation of state parameter $\tw$   \cite{Martin:2006rs,Martin:2010kz}. In section \ref{sec:inflationarypredictions} we collect predictions for the perturbation spectra of three general classes of single field models: single field, natural, and hilltop or inflection point inflation. We explore the parameter uncertainty and model degeneracy induced by the unknown post inflationary thermal history. In section \ref{sec:obs}, we explore the constraints that may be placed on the tilt and the running of the scalar spectrum by combinations of future astrophysical observations. We focus on observations of the cosmic microwave background, galaxy surveys and cosmic tomography using the 21 cm line of neutral hydrogen. In section \ref{sec:reheatconstrain} we employ the Fisher matrix formalism to forecast constraints on the post inflationary parameters for a given inflationary model. Finally, we conclude in section \ref{sec:discussion}.
 
\section{Inflation, The Matching Equation, and the Primordial Power Spectrum}\label{inflandpower}
\subsection{Equations of Motion and Slow Roll Expansion}
We begin by gathering results for the dynamics and observable properties of single field inflation.  In a flat Friedmann-Robertson-Walker universe dominated by a scalar inflaton $\phi$ with potential $V(\phi)$, the Einstein and Klein-Gordon equations are
\begin{eqnarray}
&&H^{2}  = \frac{1}{3 \Mpl^{2}}\left[\frac{1}{2}\dot{\phi}^{2}+V(\phi)\right] \, ,\\
&&  \ddot{\phi}+3H\dot{\phi}  + \frac{\ud V}{\ud\phi} =0  \, .
\end{eqnarray}
Variables have their conventional definitions and we use the reduced Planck mass, suppressing factors of $\pi$ throughout our analysis.  

We work in the potential slow roll (PSR) expansion -- this is less elegant than the Hubble slow roll (HSR) expansion  \cite{Liddle:1994dx}, but expresses physical observables directly in terms of the potential.  The first three slow roll parameters are 
\begin{eqnarray}
\epsilon(\phi) & = & \frac{\Mpl^{2}}{2}\left[\frac{V'}{V}\right]^{2} \, ,\\
\eta(\phi) & = & \Mpl^{2}\frac{V''}{V}\, ,\\
\xi(\phi) & = & \Mpl^{4}\frac{V'V'''}{V^{2}} \, .
\end{eqnarray}
During slow roll, $\epsilon \ll 1$ and $|\eta|\ll 1$ while inflation ends when $\epsilon(\phi_{end}) \approx 1$. This relationship is exact in the HSR formalism, and a good approximation in PSR.

\subsection{The Perturbations}
  
Quantum fluctuations of the metric and matter fields during inflation give rise to both curvature and density perturbations.   These fluctuations are characterized by the gauge invariant variable $\zeta$, which measures the density fluctuations on hyper-surfaces with uniform curvature. The power spectrum of $\zeta$ can be solved in the adiabatic regime ($\epsilon, \eta \approx {\rm const.}$) as an expansion about the exact power law solution \cite{Stewart:1993bc}
\begin{eqnarray}
\mathcal{P}_{\zeta}(k) & = & \frac{1}{12\pi^{2}M_{\rm pl}^{6}}\frac{V^{3}}{V'^{2}}\left[1-\left(2C+\frac{1}{6}\right)\epsilon+\left(C-\frac{1}{3}\right)\eta+ \cdots \right],
\end{eqnarray}
where $C = -2+\log 2 + \gamma\simeq -0.73$ and $\gamma$ is the Euler-Mascheroni constant. Similarly, the primordial spectrum of gravitational waves is 
\begin{eqnarray}
\mathcal{P}_{h}(k) & = & \left.[1-(C+1)\epsilon_{H}]^{2}\frac{8}{M_{\rm Pl}^{2}}\left(\frac{H}{2\pi}\right)^{2}\right|_{k = aH} ,
\end{eqnarray}
where 
\begin{eqnarray}
\epsilon_{H} & = & \epsilon\left(1-\frac{2}{3}\epsilon+\frac{1}{3}\eta\right).
%
\end{eqnarray}
 
The spectra are conventionally written as  power laws,
\begin{eqnarray}
\mathcal{P}_{\zeta}(k) & = & A_{s}(k_\star)\left(\frac{k}{k_\star}\right)^{n_{s}(k)-1} ,\, \\
\mathcal{P}_{h}(k) & = & A_{t}(k_\star)\left(\frac{k}{k_\star}\right)^{n_{t}(k)}  \, ,
\end{eqnarray}
where $n_s$ and $n_t$ are the spectral indices.  With a specified pivot $k_\star$, we measure $n_{s}(k_\star)$, $\alpha_s = \ud n_{s}(k)/\ud\log k|_{k = k_\star}$, and $r(k_\star) = \mathcal{P}_{h}(k_\star)/\mathcal{P}_{\zeta}(k_\star)$. For single field models, we have the consistency relation $n_{t} = -r/8$ (to first order in slow roll), and 
\begin{eqnarray}
\frac{1}{2}(n_{s}-1) & = & -3\epsilon+\eta - \left(\frac{5}{3}+12C\right)\epsilon^{2}+(8C-1) \epsilon\eta+
\frac{1}{3}\eta^{2} \nonumber \\ 
 &&\qquad-\left(C-\frac{1}{3}\right)\xi+\cdots,\\
%
\frac{\ud n_s}{\ud\log k} & = & 16\epsilon\eta-24\epsilon^{2}-2\xi+\cdots,\\
r & = & 16 \epsilon\left[1-\frac{2}{3}\epsilon+\frac{1}{3}\eta+2C(2\epsilon- \eta)\right] +\cdots,
\end{eqnarray}
where `$\cdots$' denotes higher-order terms in the slow-roll expansion.

\subsection{Matching Equation, Thermalization, and Expansion History}

The spectral indices and running are measured at the pivot, $k_\star$, a specific, physical scale in the present-day universe. To compare inflationary predictions to observed inhomogeneities, we  match this length to the field value at which the expressions of the previous subsection are evaluated. During inflation, a mode leaves the horizon when its wavelength matches the Hubble radius: it reenters the horizon when the post-inflationary Hubble radius expands to this (comoving) scale. Comparing a comoving scale, $k$, with the current Hubble scale we solve for $N(k)$, the number of e-folds before the end of inflation at which the mode left the horizon:
\begin{equation}
N(k) =\log\left(\frac{ H_{k}}{H_{end} }\right)+\log\left(\frac{a_{end}H_{end}}{a_{0}H_{0}}\right) -\log\left(\frac{k}{a_0 H_0} \right) \, .
\end{equation}
Assuming slow roll, we have the usual result
\begin{equation}
N(k)  \equiv  \log\left(\frac{a(t_{end})}{a(t_{k})}\right) = \int_{t_{k}}^{t_{end}}H(t)\ud t  \simeq  \frac{1}{\Mpl^{2}}\int_{\phi_{end}}^{\phi_{k}} \frac{V}{V'}\ud\phi \, .
\end{equation}
For a specific inflationary potential, the ratio $ H_{k}/H_{end}$ can be accurately evaluated but  $N(k)$ also depends on the growth of the horizon, $(aH)^{-1}$, from the end of inflation to the present day. We assume that at any given instant the post-inflationary universe has a well-defined  equation of state,  $w$, the ratio between pressure and density.\footnote{This can be an effective parameter. During coherent oscillations, $w$ rapidly oscillates between $\pm 1$, but the time-averaged value is $w=0$ in a quadratic potential. The growth of perturbations depends on the (time-dependent) ratio of the their frequency to that of the field oscillations \cite{Easther:2010mr}, but our focus here is the background dynamics.} 

Nucleosynthesis (e.g.~\cite{Steigman:2005uz}) and evidence for a cosmological neutrino background \cite{Komatsu:2010fb} imply that the universe was  in thermal equilibrium at MeV temperatures, and the evolution of the universe between this epoch and the present day is assumed to be well-understood.   The current microwave background temperature  is ${\cal {O}}( 10^{-4})$~eV, from which we infer that the universe has expanded by a factor $\sim 10^{10}$  since neutrino decoupling.  However there is no concrete evidence that the universe was thermalized at higher densities.  Conversely, as inflation ends, the universe contains few particles and is far from thermal equilibrium.  The mechanism of reheating is unknown, and usually unspecified by the inflationary potential on its own. We parameterize our ignorance by specifying  that the universe thermalizes at a temperature $T_{reh} \sim \rho_{reh}^{1/4}$.  As noted above, the only firm constraint on this parameter is $\rho_{reh}^{1/4} \gtrsim {\cal{O} }(10)$ MeV.  If inflation is a GUT scale phenomenon, the epoch between the end of inflation and the MeV scale spans a range of $10^{18}$ in energy.  Consequently,   the ratio of scales between the end of inflation and  neutrino decoupling can far exceed the separation between neutrino decoupling and the present day. Moreover, over most of this epoch the basic properties of particle physics itself are poorly understood, and amount to a ``primordial dark age''.

\begin{figure}
\includegraphics[width = 5in]{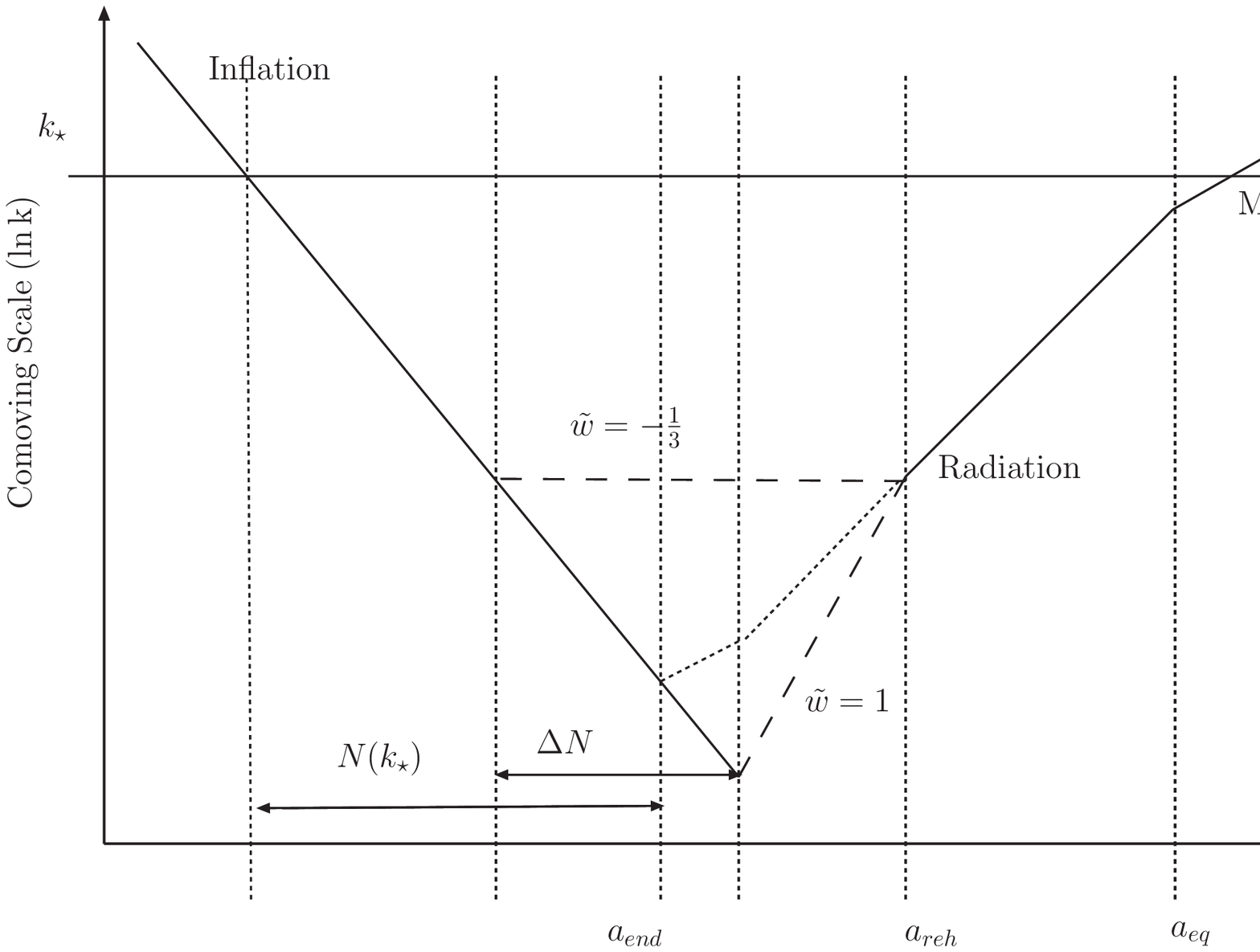}
\vspace{2mm}
\caption{The evolution of the Hubble horizon for the inflationary universe is shown in cartoon form.  The $x$-axis describes the cosmological scale factor $a(t)$ on an approximately logarithmic scale. The parameter $\tw$ describes the growth during the epoch between the end of inflation with the standard hot big bang era. Note that for smaller $\tw$  the value of $N$ at which the pivot leaves the horizon is similarly decreased. \label{fig:matching}}
\end{figure}

The ``default'' assumption  is  that the post-inflationary universe undergoes an  effectively matter dominated period of coherent oscillations, followed by thermalization, followed eventually by matter-radiation equality and the onset of structure formation.     However, a huge range of  mechanisms could interrupt this simple picture, and their impact on the expansion rate can typically be described by the equation of state $w$, the ratio of the pressure to the density. We are thus implicitly limiting ourselves to a barotropic fluid, but the discussion here could easily be generalized.  These range from  kination ($w= 1$) \cite{Chung:2007vz},  to frustrated cosmic string networks \cite{Burgess:2005sb} ($w=-1/3$), or even a short burst of  {\em thermal inflation\/} \cite{Lyth:1995ka} during which some modes would be pushed outside the horizon for a second time.   Consequently, we break the post inflationary evolution into two parts,  the unconstrained post-inflationary era, and the hot big bang era (which runs through to the present day), during which the universe is thermalized and populated with familiar standard model particles.  Considering the former, we write
\begin{equation}\label{eqn:prematch}
\log\left(\frac{a_{end}H_{end}}{a_{reh}H_{reh}}\right) = -\frac{\Delta \log (aH)^{-1}}{\Delta \log a}\Delta \log a.
\end{equation}
Rather than adding terms to equation~(\ref{eqn:prematch}) to account for each possible phase in the post-inflationary universe, we define an  effective equation of state \cite{Martin:2006rs,Martin:2010kz} 
\begin{equation}
\tilde{w} = \frac{1}{\Delta \log a} \int w(a) \ud\log a.
\end{equation}
Introducing $\tw$  replicates the growth of the scale factor \emph{and} horizon during these intermediate stages, as sketched in Figure \ref{fig:matching}.  Using standard results for the evolution of the horizon and scale factor with a fixed equation of state we derive 
\begin{equation}
\log\left(\frac{a_{end}H_{end}}{a_{reh}H_{reh}}\right) = -\frac{(1+3\tilde{w})}{6(1+\tilde{w})}\log\left(\frac{2}{3}\frac{\rho_{reh}}{V_{end}}\right) \, .
\end{equation}
The logarithmic term follows from noting that $\ddot{a} =\rho+3p =0$ at the end of inflation,  giving $\rho_{end} = 3V_{end}/2$.

Almost any conceivable evolution between the end of inflation and the present day  is now encoded in   $\tw$ and $\rho_{reh}$.   This argument relies on astrophysically relevant modes being outside of the horizon until after thermalization: modes inside the horizon respond directly to the equation of state, and need the full transfer function -- this distinction will be important for direct detection gravitational wave experiments  \cite{Boyle:2005se,Easther:2008sx}.    Assuming reheating is followed by the radiation and matter dominated epochs, one arrives at the {\em matching equation\/}:
\begin{eqnarray}\nonumber \label{eq:matchgen}
N(k) & = & 56.12-\log\left(\frac{k}{k_{\star}}\right)+\frac{1}{3(1+\tilde{w})}\log\left(\frac{2}{3}\right)+\log\left(\frac{V_k^{\frac{1}{4}}}{V^{\frac{1}{4}}_{end}}\right)\\ & & +\frac{(1-3\tilde{w})}{3(1+\tilde{w})}\log\left(\frac{\rho_{reh}^{\frac14}}{V_{end}^{\frac14}}\right)+ \log\left(\frac{V_{k}^{\frac14}}{10^{16}{\rm GeV}}\right),
\end{eqnarray}
where $k_{\star} = 0.05$ Mpc$^{-1}$ is chosen as the pivot scale in what follows.\footnote{Note that the optimal choice of pivot depends on the dataset being considered \cite{Peiris:2006sj,Cortes:2007ak}.}  Notice that all dependence on the present-day Hubble parameter, $h$, cancels. A fluid with $p> \rho$ has a superluminal sound speed, so we expect $\tw \le 1$.  Conversely,  inflation ends at $V_{end}$, so $\tw > -1/3$, to ensure that the comoving horizon does not shrink further during the (supposedly) post-inflationary evolution. The definition of $\tw$ can accommodate a  secondary burst of inflation, but the overall evolution must be such that modes are re-entering rather than leaving the horizon.  If the universe thermalized instantaneously and the number of degrees of freedom in the thermal bath does not change as the universe expands, $\tw=1/3$. Conversely, $\tw=0$ implies the universe is (on average) matter dominated until $\rho^{1/4}= \rho_{reh}^{1/4}$.

\begin{figure}
\includegraphics[scale=0.65]{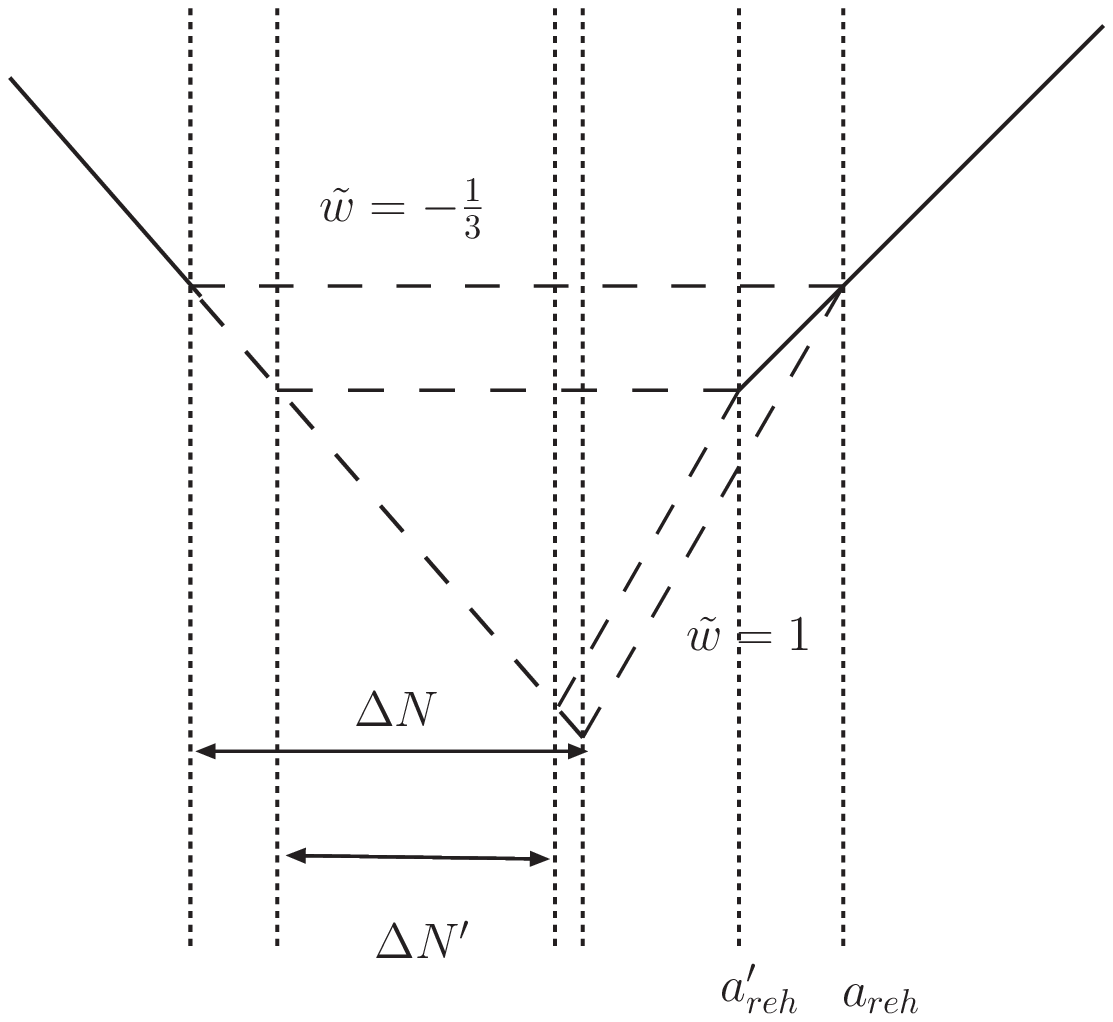}
\includegraphics[scale=0.65]{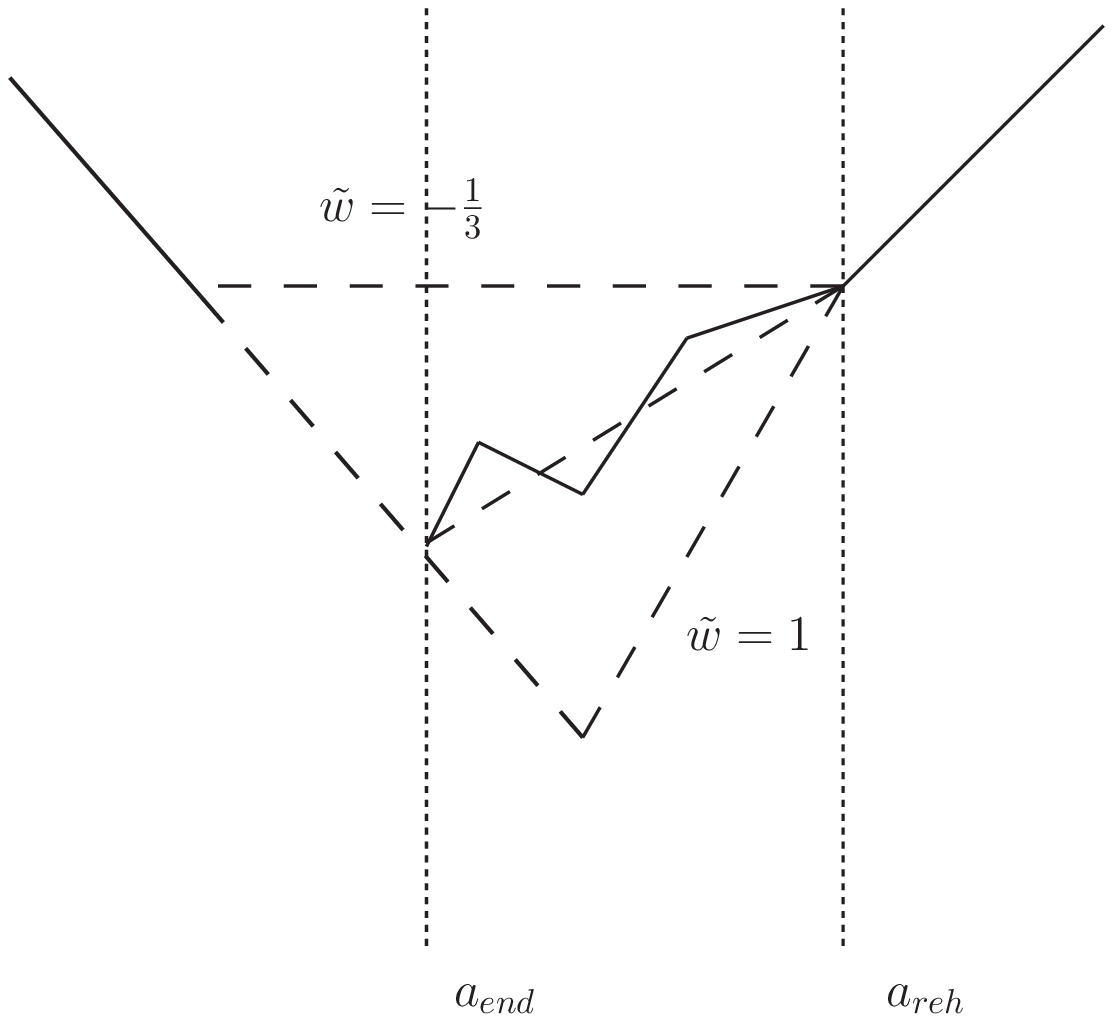}
\vspace{2mm}
\caption{ We illustrate the impact of changing our the assumed value of $\rho_{reh}$. The left hand panel demonstrates that setting a high value of  $\rho_{reh}$ reduces the total uncertainty in $N$, whereas the right hand panel  illustrates a specific (and baroque) post-inflationary history (solid line) which then fixes the effective expansion history and $\tw$.  \label{fig:matchingdetail}}
\end{figure}

Note that  $\rho_{reh}$ usually refers to the actual energy scale of thermalization, whereas in this treatment  $\rho_{reh}$ is the energy scale at which thermalization is guaranteed to have occurred. As noted above, direct experimental constraints require $\rho_{reh}^{1/4} \gtrsim {\cal{O} }(10)$ MeV.  Theoretical considerations may be used to justify a much higher value of $\rho_{reh}^{1/4}$, but it is important to draw a clear distinction between a theoretical prior and an observational bound.  As we illustrate in Figure~\ref{fig:matchingdetail}, the value of $\tw$ is a function of $\rho_{reh}$, given a fixed cosmological evolution.

 Equation (\ref{eq:matchgen}) appears in the literature in various forms.    Assuming a matter dominated phase followed by thermalization ($\tw= 0$)  one can derive the result quoted by \cite{Liddle:2003as,Adshead:2008vn} 
\begin{eqnarray}
N(k) &=& 55.75 -
\log\left[ \frac{k}{k_{\star}}\right]-\log\left[\frac{10^{16}\textrm{GeV}}{V_{k}^{\frac{1}{4}}}\right]+\log\left[\frac{V_{k}^{\frac{1}{4}}}{V_{end}^{\frac{1}{4}}}\right] \nonumber \\
&&
-\frac{1}{3}\log\left[\frac{V_{end}^{\frac{1}{4}}}{\rho_{reh}^{\frac{1}{4}}}\right]  \, .
\end{eqnarray}
where $\rho_{reh}$ is again the usual thermalization scale.     This expression assumes the effective  number of thermal degrees of freedom is constant as the universe expands, and makes no mention of the contribution of neutrino mass or dark energy to the expansion of the  universe after nucleosynthesis.

We distinguish three {\em priors\/} for the evolution of the post-inflationary universe. The first, and most restrictive, is to set $\tw=1/3$ and stipulate that the universe is thermalized at all scales between the end of inflation and matter-radiation equality.  Secondly, we can posit a matter dominated phase, followed by thermalization, so $\tw =0$, and  $\rho_{reh}$ is a free parameter. Finally, if we foreswear all knowledge of the post-inflationary expansion history that is not astrophysically verified, $\tw$ is a free parameter with $-1/3 \le \tw \le 1$.

It is useful to view $\Delta N$ as a ``shift'' in the pivot $k_\star$, which is mapped to  $k=k_\star e^{\Delta N}$, where $k$ is measured in a universe which thermalized immediately after inflation.    
Even with the ``middle'' assumption of matter domination followed by thermalization,  $\Delta N(k) \sim -9$ if $(\rho_{end} /\rho_{reh} )^{1/4} \sim 10^{12}$, relative to instant reheating.  Anticipating the results of the following section, simple inflationary models have  $10^{-4} \lesssim |\alpha_s| \lesssim 10^{-3}$. Recalling the definition $\alpha_s = d n_s/d\log{k}$,   the corresponding  {\em theoretical\/}  uncertainty in $n_s$ is (conservatively) $ |\delta n_s| \sim 5 \times 10^{-3}$, without any appeal to exotic dynamics in the early universe. For many inflationary models this theoretical ambiguity can easily exceed the  forecast {\em statistical\/} uncertainty in $n_s$ expected from Planck \cite{Planck:2006uk}.   Finally, with the full range of $\tw$, equation (\ref{eq:matchgen}) gives $-25 <  \Delta N < 8.4$ if we insist on thermalization at the TeV scale, and an even greater range if we set   $\rho_{reh}$ at the MeV scale. 
 
\section{Inflationary Predictions}\label{sec:inflationarypredictions}

We now study the connection between  $\tw$ and the predictions of specific  inflationary models.  Given a candidate model of particle physics which makes a full (and computable) set of predictions for the evolution of the universe,  $\tw$ would not be a free parameter. In practice, however,  inflationary models are specified independently of a full theory of particle physics.   Consequently, the predictions of inflationary models should properly be considered as a combination of the inflationary dynamics and $\tw$.    We begin by  collating the predictions for $r$, $n_s$ and $\alpha_s$ of several broad classes of simple two parameter inflationary models.   In each case, we eliminate one free parameter from the model by matching the amplitude of the perturbation spectrum at the pivot to the observed value, $A_s$ in the current concordance cosmology. In practice, the following results depend very weakly on the precise value of this parameter.

\subsection{Single term potentials: $\phi^n$ }

We start with models whose potentials are simple powers of $\phi$, 
\begin{equation}
 V = \lambda M_{Pl}^{4-n} \frac{\phi^{n}}{n} ,
 \end{equation}
 so $\lambda$ is a dimensionless constant.    Once upon a time,  one might have assumed that $n$ is an even integer, for which $n=2$ is currently the only viable value.  However, recent developments in string cosmology yield models   with $n=2/5$, $n=2/3$ and $n=1$, so we treat $n$ as a continuous quantity  \cite{McAllister:2008hb,Flauger:2009ab}.  In these models, inflation ends at $\phi_{end} = n \Mpl/\sqrt{2}$ and the field value $N$ e-folds before the end of inflation, $\phi_{N}$, is
\begin{equation}
 \phi_{N}  =  \sqrt{n \left(2N+\frac{n}{2}\right)}\Mpl \, .
\end{equation}
We  can write the slow roll parameters as a function of $N$,
\begin{equation}
\epsilon(N)  =   \frac{n}{\left(4N+n\right)}, \, \quad
\eta(N)  =  \frac{2(n-1)}{\left(4N+n\right)}, \, \quad
\xi(N)  = \frac{4(n-1)(n-2)}{\left(4N+n\right)^{2}},
\end{equation}
so the spectral parameters are (at lowest order in $N$)
\begin{equation}
n_{s}  =  1-\frac{2n+4}{\left(4N+n\right)}, \, \quad
r  =  \frac{16n}{4N+n}, \quad
\alpha_s  =-\frac{8(2+n)}{(4N+n)^{2}}.
\end{equation}
Note that  $r \rightarrow 0$ as $n \rightarrow 0$, but $\alpha_s \rightarrow -1/N^2$ in the same limit.  Given that $N$ is bounded above,  the running remains non-zero  in these models, even as the tensor component becomes vanishingly small.

\subsection{Natural Inflation }

Natural inflation \cite{Freese:1990rb} is governed by the axion-motivated potential,
\begin{equation}
V(\phi) =  \Lambda^{4}\left[1+\cos\left(\frac{\phi}{f}\right)\right].
\end{equation}
Setting $\epsilon = 1$ we solve for the end-point of inflation, 
\begin{equation}
  \cos\left(\frac{\phi_{end}}{f}\right)  = \frac{1-\frac{2f^{2}}{M_{\rm Pl}^{2}}}{1+\frac{2f^{2}}{M_{\rm Pl}^{2}}}    \, .
\end{equation}
Making no assumptions about $f/M_{\rm Pl}$ one finds
\begin{equation} 
N  =   -\frac{f^{2}}{M_{\rm Pl}^{2}} \log\left[\frac{1-\cos\left(\frac{\phi}{f}\right)}{1-\cos\left(\frac{\phi_{end}}{f}\right)}\right]  \, .
\end{equation}
 When $f$ is small relative to $M_{\rm Pl}$, inflation ends when $\cos(\phi/f)$ is close to unity -- in other words, at the top of the hill. Conversely, when $f$ is substantially larger than  $M_{\rm Pl}$, inflation continues until the field is close to its minimum and $\cos{(\phi_{end}/f)} \rightarrow -1$. In this case the final 60 e-folds of inflation are generated as $\phi$  approaches the bottom of the axion potential: $\epsilon$, $\eta$ and $\xi$ are all positive, yielding an observable tensor spectrum and a moderate running.  Conversely, for smaller $f$, astrophysically relevant modes are generated at smaller values of $\phi$.  For extremely small $f$, $\epsilon$, and $\xi \ll |\eta|$ in this limit, and we have a vanishing running, a vanishing tensor amplitude, and $n_s-1$ is fixed by $\eta$. The  lower bound on $n_s$ sets the minimal value of $f$:  the  three year WMAP data set gives $f \gtrsim 0.7 \sqrt{8 \pi}  M_{\rm Pl} $ or $f \gtrsim 3.5  M_{\rm Pl} $  \cite{Savage:2006tr}. Interpolating between these limits suggests that $\alpha_s(k_\star)$  approaches zero as $f$ decreases, and this expectation is confirmed in the numerical analysis that follows.

\begin{figure}[tbp]
 \includegraphics[width = 15cm]{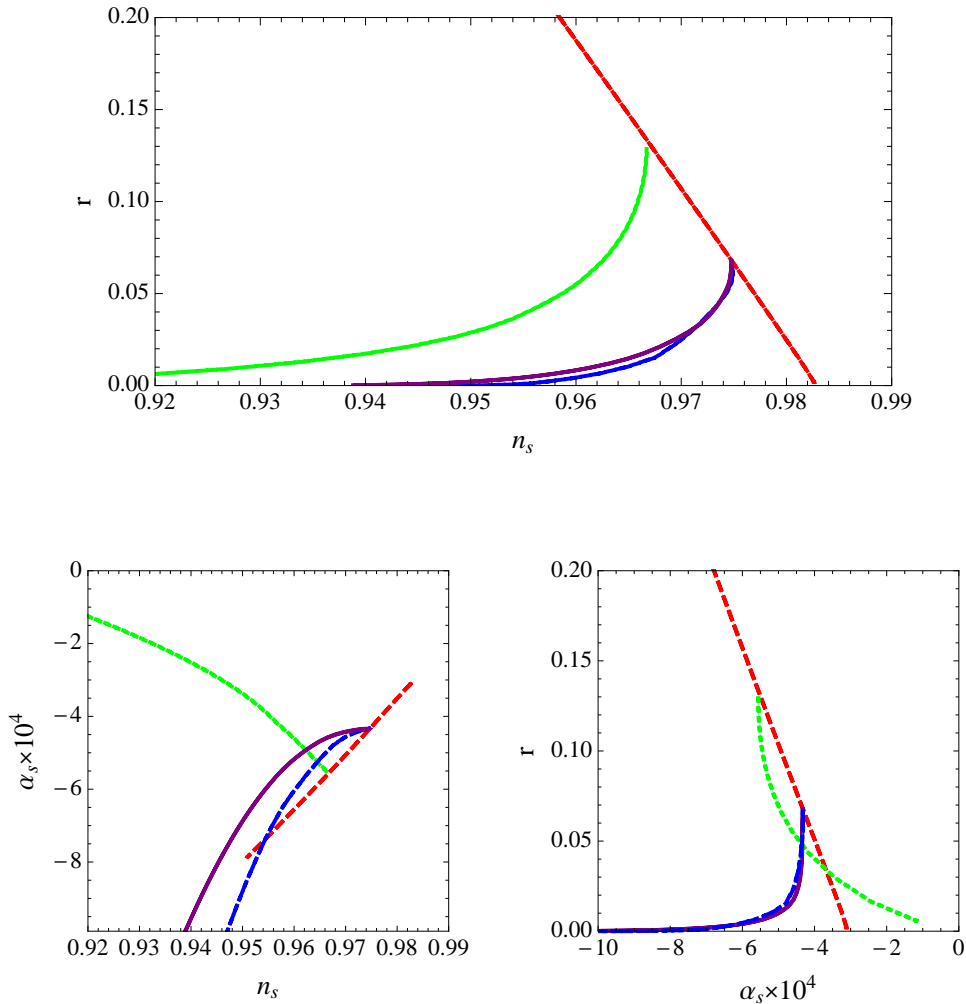} 
\caption{\label{fig:instant} Top panel: $n_s$ vs. $r$, Bottom, $n_s$ vs. $\alpha_s$, $\alpha_s$ vs. $r$. We plot $\phi^n$ inflation (red), natural inflation (green),   inflection (purple) and hilltop (blue).  Hilltop and inflection point inflation meet the $\phi^n$ curve at the point where $n=1$;  natural inflation is degenerate with $m^2 \phi^2$ inflation in the limit that $f$ is very large.}
\end{figure}

 \begin{figure}[tbp]
\includegraphics[width = 16cm]{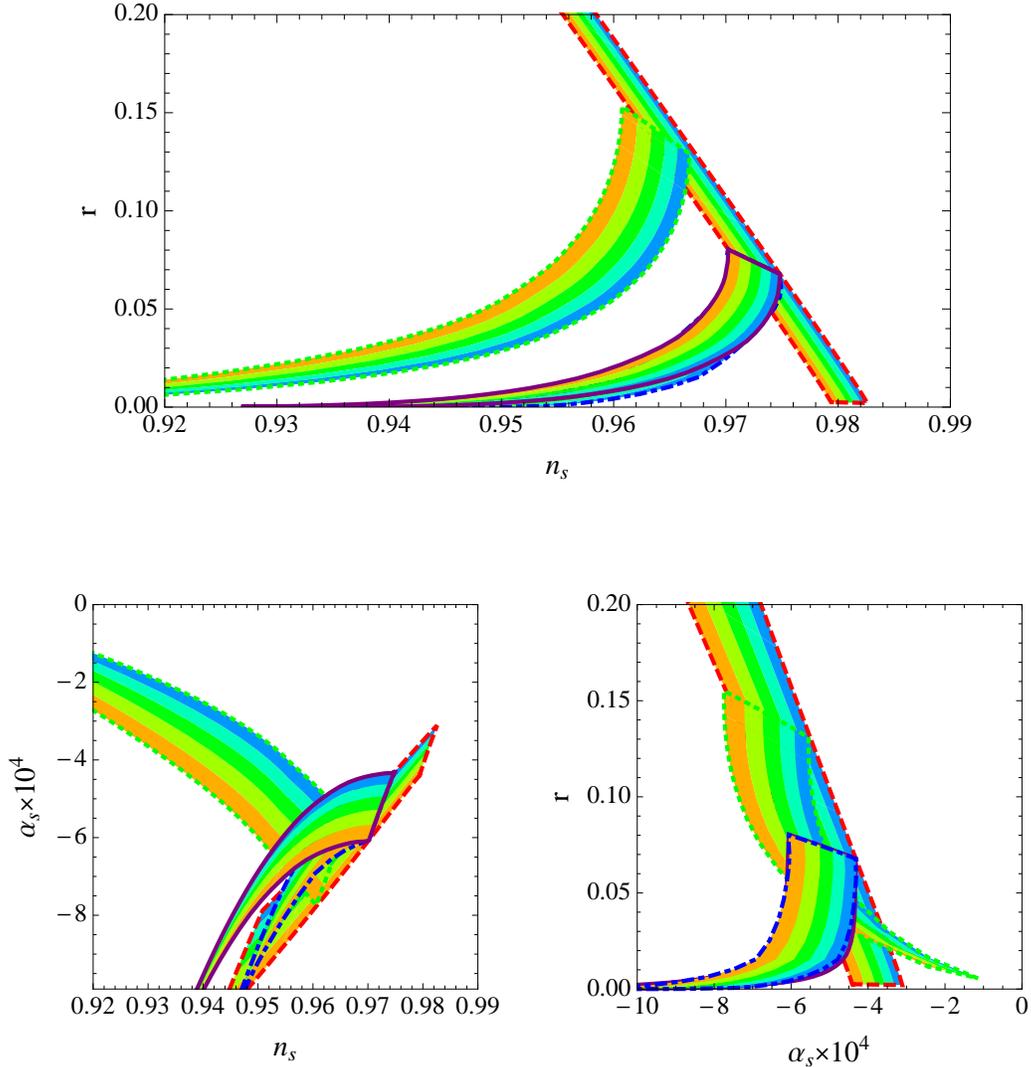}
\caption{\label{fig:general} Same models as    Figure~\ref{fig:instant}, assuming a period of matter dominated expansion before thermalization, with $T_{reh} >10^3$ GeV.  Colors denote the (logarithmic) reheating temperature -- orange is low, and blue is high.   }
\end{figure}

\subsection{Hilltop and Inflection Point Inflation}
 
Now consider the potential
\begin{equation}
V(\phi) = \Lambda^4 - \lambda M_{\rm Pl}^{4-n}  \frac{\phi^n }{n} \, ,
\end{equation} 
 where $\lambda$ is a dimensionless constant. This potential can be specialized to several distinct models: e.g. hilltop ($n=2$ and $n=4$) and inflection point ($n=3$).     Like natural inflation, these models have two distinct limits.   In one limit $\Lambda$ is relatively small, and almost all inflationary growth occurs near $\phi =0$, with $\epsilon \ll \eta$.  Alternatively, if $\Lambda^4$ is large (relative to the GUT scale), the physically relevant portion of inflation occurs with $\phi$ far from the origin, and $\epsilon \approx |\eta|$. In the latter case the potential is effectively linear, and    $V(\phi) \propto \phi$ -- or $\phi^n$, with $n=1$.   If we restrict attention to ``small $\phi$" we find
\begin{equation}
N \approx \frac{1}{(n-2)\Mpl^{6-n}} \frac{\Lambda^4}{\lambda} \frac{1}{\phi^{n-2}}.
\end{equation} 
To lowest order we can compute
\begin{equation}
P_\zeta = \frac{1}{12 \pi^2 \Mpl^{14-2n}} \frac{\Lambda^{12}}{\lambda^2} 
\left[ \frac{ (n-2) \lambda N}{\Mpl^{n-6} \Lambda^4}\right]^{\frac{(2n -2)}{(n-2)}}.
\end{equation}

When $n=4$, $\Lambda$ drops out of the above expression and inflation can occur at any energy.\footnote{To be strictly accurate, the energy scale enters logarithmically via $N$ when we solve the constraints self-consistently.} Consequently, for $n=4$,  $r$ is essentially a free parameter.  With $n=3$, we recover the inflection point model \cite{Baumann:2007np}. Current data suggests   $n_s< 1$, so we can ignore the singularity in the slow roll expressions at the $\phi=0$, since red spectra require $\phi>0$.  For these models, as $\epsilon$  becomes small, $\xi$ grows large. Consequently, in the low $r$ limit we find a substantial, negative running for the hilltop and inflection scenarios.

\subsection{Observables and the Post-Inflationary Thermal History}

\begin{figure}[tbp]
\includegraphics[width = 15cm]{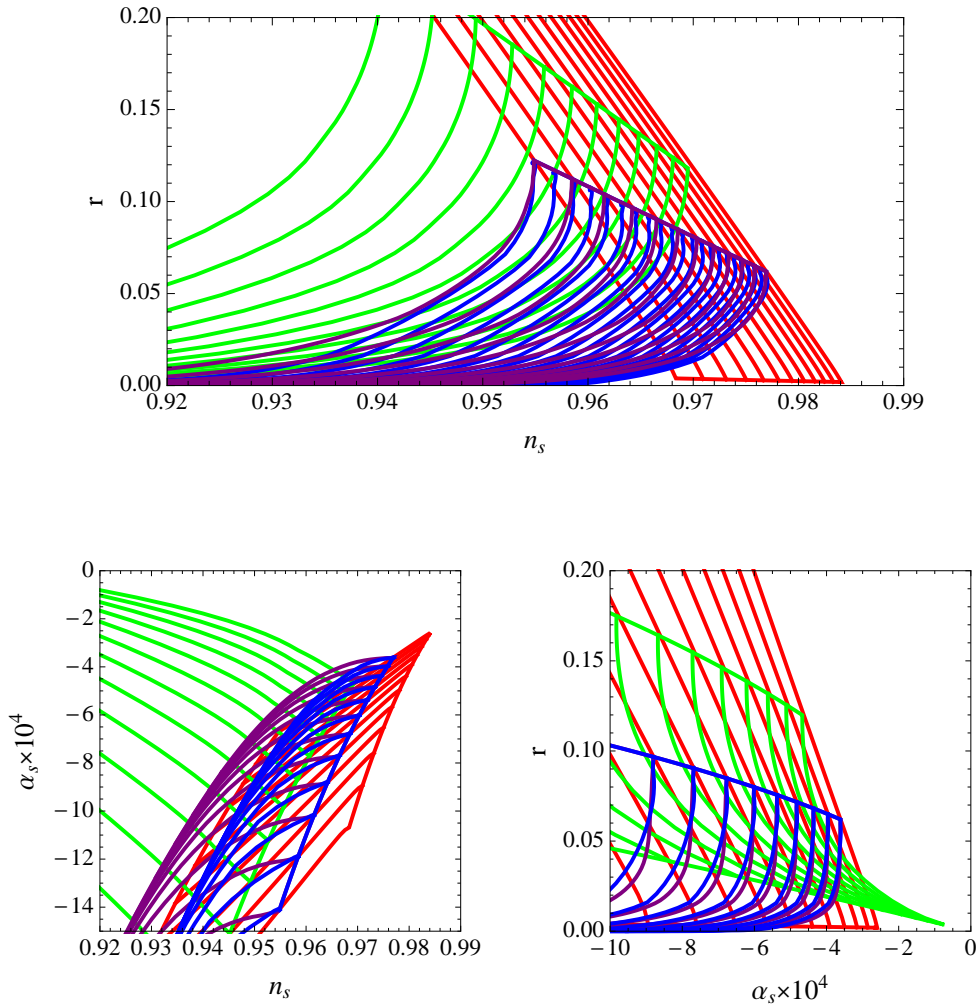}
\caption{\label{fig:full} Same models as    Figures~\ref{fig:instant} and \ref{fig:general}, but for the full range of $\tw$ and $\rho_{reh}^{1/4} = 1$~TeV.  Parallel lines for each model denote fixed values of $N$.  }
\end{figure}  

The above catalogue made no direct reference to the post-inflationary evolution of the universe.   Figure~\ref{fig:instant}  shows the parameter values derived assuming instant thermalization at the end of  inflation. Each model has two free parameters to the potential, one of which is removed by matching to the observed amplitude of the perturbations at CMB scales. The form of these plots is very weakly sensitive to the precise value we take for this parameter, so these plots define one-dimensional curves in the parameter space.  

If we assume that thermalization is preceded by a matter dominated phase, and that reheating has occurred before the universe reaches TeV scale densities,  $0> \Delta N \gtrsim -9$, relative to the instant reheating case.    Figure~\ref{fig:general} replots the models from Figure~\ref{fig:instant} including the effect of the unknown thermalization scale. A nontrivial portion of the parameter space is now occupied, and each parameter plane contains degeneracies which are not present for instant preheating.     Finally,  Figure~\ref{fig:full} plots the inflationary observables with $\rho_{reh}^{1/4}$ at the TeV scale and arbitrary $\tw$. In the absence of a theoretical prior for the post-inflationary evolution   these simple models are significantly degenerate, and cover a large fraction of each parameter plane.

\begin{figure}[tbp]
\includegraphics[width = 7cm]{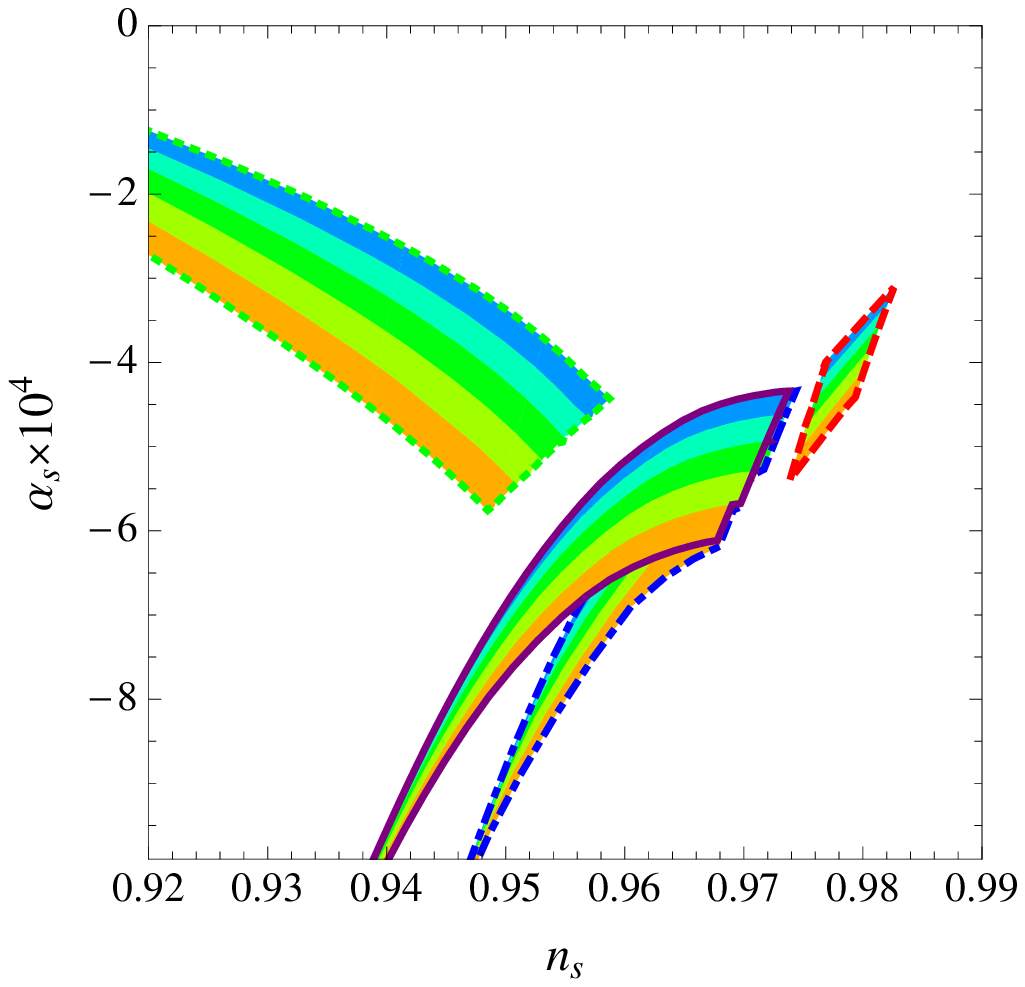} \includegraphics[width = 7cm]{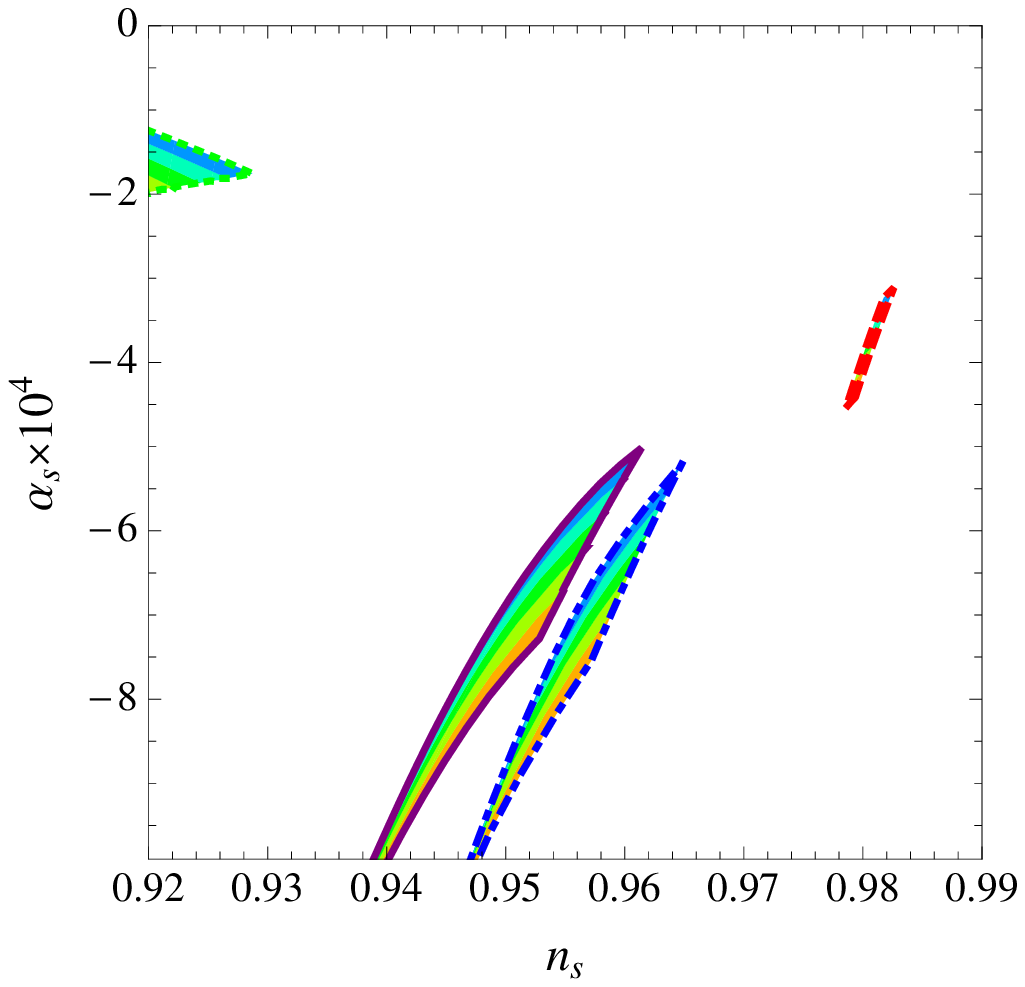}
\caption{ \label{fig:rcuts} We now plot the ranges of $n_s$ and $\alpha_s$ given by our set of models when $r<0.05$ (left) and $r<0.01$ (right).  }
\end{figure}

We can also ask how these plots change when we have more information about some of these parameter values.   For instance, assuming that Planck and the next generation of sub-orbital polarization experiments put a tight upper bound on $r$: Figure~\ref{fig:rcuts} shows the permitted regions of the $(n_s,\alpha_s)$ plane for our models that satisfy $r<0.05$ and $r<0.01$.    Further, in either of these scenarios $n_s$ is also likely to be tightly constrained, to the point that at least one of the classes of models we consider will be ruled out.

For this collection of models, $r$ varies by orders of magnitude, but typically $-10^{-3} \lesssim \alpha_s  \lesssim -10^{-4}$, so while $\alpha_s$ is small, it is far more consistent. Detecting a running at the lower end of this range will be a challenging task as we see in the next section, but this range provides a fairly clear target for future observational campaigns.

\section{Observational Constraints }
\label{sec:obs}

In this section, we explore how we might probe the running at the $10^{-3}\gtrsim |\alpha_s| \gtrsim10^{-4}$ level.    As a base level we estimate the constraints from Planck \cite{Planck:2006uk} using the Fisher matrix formalism \cite{Kamionkowski:1996ks} and using instrumental parameters taken from \cite{Baumann:2008aq}. These calculations suggest that Planck should achieve constraints on the tilt and running of $\delta n_s=0.003$ and $\delta\alpha_s=0.005$ respectively.  To improve upon this will require information from large scale structure, which may come from galaxy surveys, 21 cm experiments, cosmic shear surveys, or the Lyman alpha forest.  We will consider only the first two of these probes here as having the most potential to provide constraints on models of slow roll inflation.

Cosmic shear surveys will be very powerful in constraining dark energy models, but provide only limited improvement to constraints on inflationary parameters \cite{Ishak:2003zw}.   Perhaps a factor of two improvement on $n_S$ and $\alpha_S$ beyond constraints from Planck level data seem possible \cite{Song:2004tg,Kitching:2008dp}.  Several attempts have been made to use the Lyman alpha forest to constrain inflation using spectra from SDSS \cite{viel2004,mcdonald2005,Seljak:2006bg}.  While this could be very powerful since the forest constrains scales $k\approx1\iMpc$, greatly extending the lever arm, modeling the ionization state and thermodynamic properties of the intergalactic medium to convert flux measurements into a density power spectrum is very challenging \cite{Zaldarriaga:2000mz}.  The Baryon Oscillation Sky Survey (BOSS) is expected to increase the power of Lyman alpha forest constraints by greatly increasing the number of quasar lines of sight and allowing non-parametric reconstruction of the inflationary power spectrum \cite{Bird:2010mp}.  Robust estimation of inflationary constraints from the Lyman alpha forest is an important topic, but it lies beyond the scope of this paper and we leave it to future work.

\subsection{Galaxy surveys}
\label{sec:galaxies}

\subsubsection{Fisher matrix formalism}

Cosmological constraints from galaxy clustering and  baryon acoustic oscillations (BAO) play a crucial role in breaking degeneracies in the CMB \cite{EHT99,colombo2009}.  The constraining power of a galaxy survey is determined by three main factors: the volume of sky surveyed $V_{\rm survey}$, which determines the number of $\mathbf{k}$ modes that fit within the survey; the number density of galaxies observed $n_{\rm gal}$, which determines the importance of shot noise; and the maximum wavenumber where the analysis ends $k_{\rm max}$, usually set by the scale on which structure becomes non-linear.  Since the non-linear scale is a strong function of redshift, we must also consider the redshift range $(z_{\rm min},z_{\rm max})$ of the survey.  It has been shown \cite{takada2006} that high redshift galaxy surveys can reasonably constrain the running at the level $\alpha_s\approx10^{-3}$.  We explore both this scenario and more ambitious hypothetical examples to probe the limits of the possible.

Following the basic formalism set forward in \cite{EHT99}, we calculate the constraints for Planck in combination with galaxy surveys.  The Fisher matrix for a galaxy survey can be written as \cite{EHT99}
\begin{equation}\label{fisher_gal}
F_{ij}=\int_0^{k_{\rm max}} \frac{\ud^3\mathbf{k}}{2(2\pi)^3}\frac{\partial\log P(\mathbf{k})}{\partial p_i}\frac{\partial\log P(\mathbf{k})}{\partial p_j}\left[\frac{n_{\rm gal}P(k,\mu)}{n_{\rm gal}P(k,\mu)+1}\right]^2V_{\rm{survey}},
\end{equation}
where the derivatives are evaluated using the cosmological parameters of the fiducial model.  The $1-\sigma$ errors on the parameter $p_{i}$ are then given by $\Delta p_i=\sqrt{F^{-1}_{ii}}$.  

In calculating the power spectrum of galaxies, we allow for  biasing and redshift space distortions, so that $P(k,\mu)=(1+\beta\mu^2)^2P_{g}(k)+P_0$, where $\mu=k_{||}/k$, $P_0$ is residual shot noise, and the distortion parameter $\beta=\Omega_m(z)^{0.6}/b_1$.  This gives the intrinsic power spectrum, and the $k$ space power spectrum inferred from angular and redshift measurements is further modified by the Alcock-Paczynski effect \cite{alcock1979}.  We assume that on large scales the galaxies are linearly biased with respect to the dark matter distribution so that $P_g(k)=b_1^2P_\delta(k)$, with linear bias $b_1$.  Nonlinear biasing can introduce considerable systematic uncertainty into the galaxy power spectrum on small scales, and  we discuss this in more detail below.

In addition to the non-linearity of galaxy bias, we should worry about the non-linear evolution of the underlying density field which can erase cosmological information on small scales.  We initially model the cut-off scale $k_{\rm max}$ using the prescription of \cite{se2003probe}, who assume that non-linear effects are important for $k>k_{nl}=\pi/(2R_{nl})$, where $R_{nl}$ is the scale on which averaged density fluctuations $\sigma(R_{nl})=0.5$.  Beyond $k_{\rm max}$ we are throwing away cosmological information, which a better understanding of non-linear effects (e.g. via higher order perturbation theory) might allow us to exploit. Consequently, we generalize this definition to $k_{nl}(\gamma)$, where $\sigma(R_{nl})=\gamma$, and use $\gamma=0.5$ unless otherwise specified.

In calculating the Fisher matrix, we include the cosmological parameters ($\Omega_m h^2$, $\Omega_b h^2$, $\Omega_\Lambda$, $A_S$, $\tau$, $n_s$, $\alpha_s$, $r$).  To this we add parameters describing the galaxy bias, $b_1$, redshift-space distortions, $\beta$, and shot noise, $P_0$, in each redshift bin.  The fiducial value for the bias is calculated using the Sheth-Torman peak-background split \cite{sheth1999} appropriate for dark matter halos and assuming that each halo hosts only a single galaxy (see \cite{cooray2002} for a review of the technical details).  We set the minimum halo mass $M_{\rm min}$ using the mean number density of galaxies in the survey $n_{\rm gal}$ by assuming that the number density of halos $n_{\rm halo}(M>M_{\rm min})=n_{\rm gal}$.  This is appropriate, for example, if the survey collects only the brightest object within the survey volume.  

Our numerical calculations and forecasts use the fiducial cosmology $\Omega_m=0.3$, $\Omega_\Lambda=0.7$, $\Omega_b=0.046$, $H=100h\,\rm{km\,s^{-1}\,Mpc^{-1}}$ (with $h=0.7$), $n_S=0.95$, $\tau=0.1$, and $\sigma_8=0.9$, consistent with the  WMAP seven year data \cite{komatsu2009}.  We note that our fiducial set of cosmological parameters is not exhaustive and could be expanded, for example to include the dark-energy equation of state $w$ and the sum of neutrino masses $M_\nu$ as free parameters.  Including more parameters will degrade the constraints somewhat beyond those considered here.

\subsubsection{Constraints from  galaxy surveys}

We first delineate the parameter space accessible to galaxy surveys to get a sense of how well present and future galaxy surveys might perform.  We initially focus on the $\Omega-n_{\rm gal}$ plane, where $\Omega$ is the observing area, noting that there is a tension between going deep, to increase $n_{\rm gal}$, or wide, to increase $\Omega$.  As a point of reference, the HUDF [Hubble Ultra Deep Field], which covers only 11 sq. arcmin., has a number density of galaxies $n_{\rm gal}=0.02 \iDens$ at $z=5$ \cite{bouwens2007}, obtained from integrating the fitted luminosity function of drop outs down to the observational magnitude limit of $M=-16$. This gives a sense of the upper limit for the number density of galaxies accessible to a very futuristic survey.  To a similar brightness limit, about twice this number of Lyman break galaxies are observed at $z=3$ \cite{steidel1999}.

\begin{figure}[tbp]
\begin{center}
\includegraphics[scale=0.85]{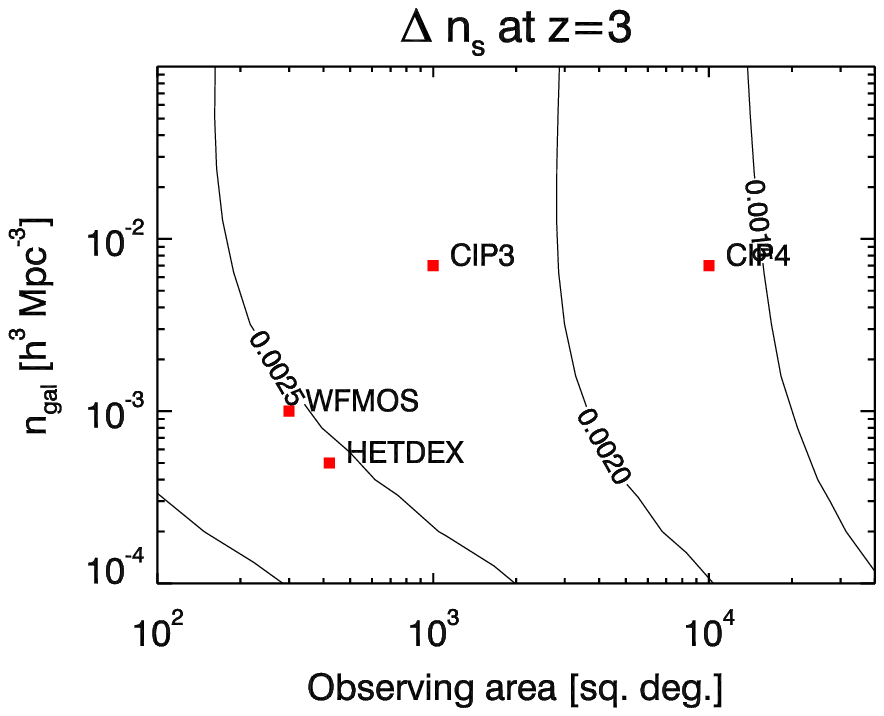}
\includegraphics[scale=0.85]{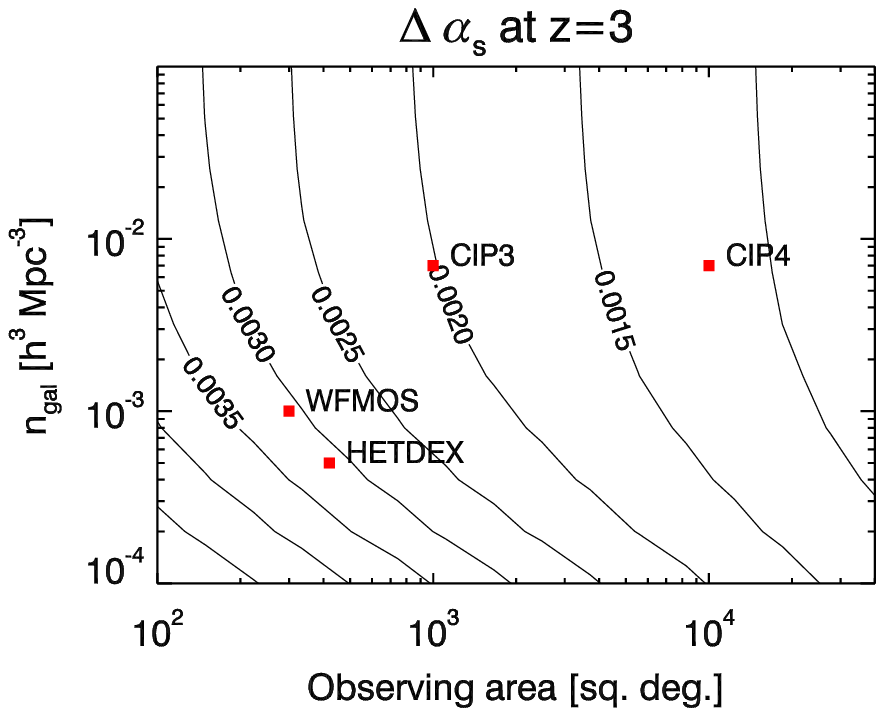}
\caption{Contour plot of $\Delta n_s$ (left panel) and $\Delta\alpha_s$ (right panel) at the 1-$\sigma$ level in the $\Omega-n_{\rm gal}$ plane at $z=3$ for a galaxy survey in combination with Planck.}
\label{fig:nscal_area_ngal_plot}
\end{center}
\end{figure}

We show the constraints on $n_s$ and $\alpha_s$ in Figure \ref{fig:nscal_area_ngal_plot} for different values of $(\Omega,\,n_{\rm gal})$ for a galaxy survey centered at $z=3$ of width $\Delta z=1$, covering an angular region $\Omega$.  Increasing $\Omega$ always improves the constraints, since more volume gives greater statistical information.  Increasing $n_{\rm gal}$ also improves the constraints, but only up to the point that shot-noise becomes negligible.  Galaxy surveys have a larger impact on $\alpha_s$ than $n_s$,  since $n_s$ is already tightly constrained from CMB  data alone.   To give a sense of what future surveys might achieve,  Figure \ref{fig:nscal_area_ngal_plot} includes points denoting the expected performance of several upcoming or proposed spectroscopic galaxy surveys that will reach $z=3$, including  WFMOS with $(\Omega,n_{\rm gal})=(300,10^{-3})$ (in units of (arcmin$^2$, $h^3{\rm Mpc}^{-3}$)), and HETDEX with $(\Omega,n_{\rm gal})=(420,0.5\times10^{-3})$. Moreover, we present results for two possible configurations of the {\em Cosmic Inflation Probe} (CIP)\footnote{http://www.cfa.harvard.edu/cip/}, a proposed satellite mission designed to observe H$\alpha$ galaxies over the range $z=1.9-6.6$, CIP3 with $(\Omega,n_{\rm gal})=(10^3,7\times10^{-3})$  and a somewhat beefed up version, CIP4 with $(\Omega,n_{\rm gal})=(10^4,1.5\times10^{-3})$. CIP is designed to reach higher in redshift to $z\lesssim6.5$, which would further improve its inflationary constraints. 

\begin{figure}[tbp]
\begin{center}
\includegraphics[scale=0.4]{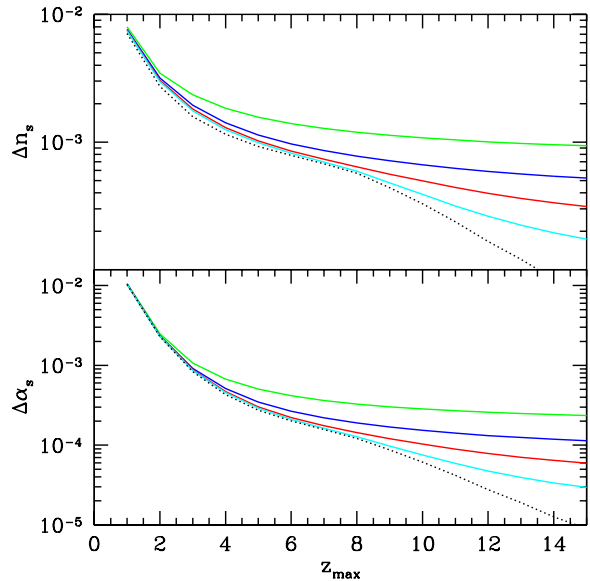}
\caption{Constraints on inflationary parameters for Planck in combination with a full sky survey up to redshift $z_{\rm max}$ for $n_{\rm gal}=10^{-4}-10^{-1}\,h^3{\rm Mpc^{-3}}$.  Black dotted curve is the cosmic variance limit.}
\label{fig:zloop_full}
\end{center}
\end{figure}

Extra volume can be added by expanding the redshift range covered by the survey.  In Figure \ref{fig:zloop_full}, we show the parameter constraints for an all sky galaxy survey in which we add redshift bins of width $\Delta z=1$ up to a maximum redshift $z_{\rm max}$.  Observations at higher redshift benefit from probing both more volume (for fixed $\Omega$), and because shorter comoving scales are still in the linear regime, leading to tighter bounds on $n_s$ and $\alpha_s$.  However, the extra linear scales are only useful if shot noise is not significant, as can be seen in the saturation of the curves in Figure \ref{fig:zloop_full} for fixed $n_{\rm gal}$.  In this calculation, the constraint is usually dominated by the highest redshift bin.

It is apparent that galaxy surveys focussed on redshifts $z\lesssim3$, such as SKA or JDEM, will not detect running at levels below $|\alpha_s| \sim10^{-3}$.  Pushing into the range $\Delta\alpha_s={\rm few}\,\times10^{-4}$, characteristic of the slow roll models  considered here requires reaching higher redshifts.  Although these might be accessible to a space based survey, attaining such redshifts may require different techniques, such as 21 cm experiments.  

\begin{figure}[tbp]
\begin{center}
\includegraphics[scale=0.85]{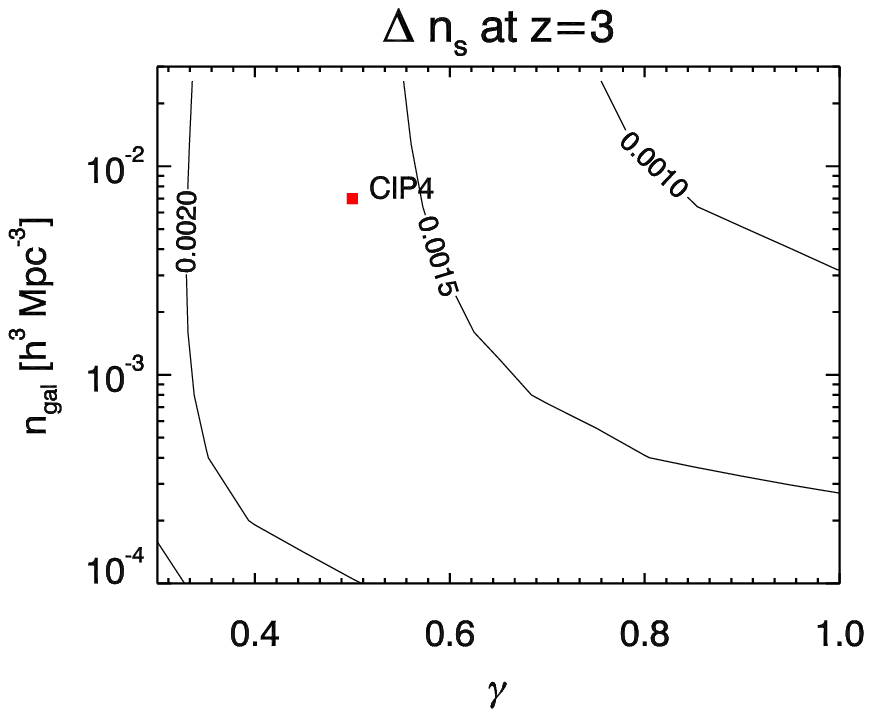}
\includegraphics[scale=0.85]{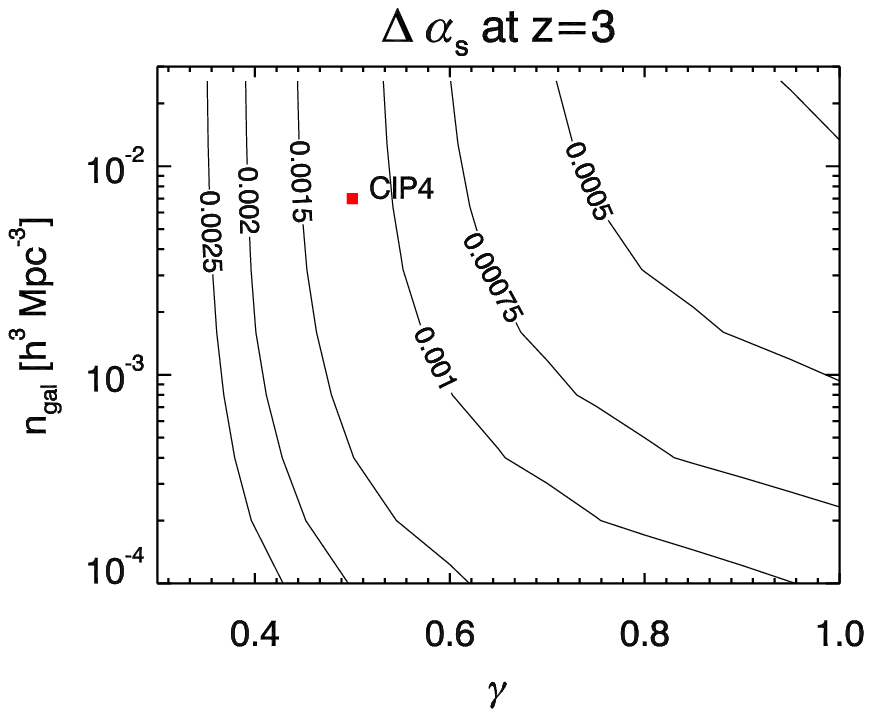}
\caption{Contour plot of $\Delta n_s$ (left panel) and $\Delta\alpha_s$ (right panel) in the $\gamma-n_{\rm gal}$ plane at $z=3$ for $\Omega=10,000\,{\rm sq.\,deg.}$.}
\label{fig:kmax_ngal_plot}
\end{center}
\end{figure}

Alternatively, rather than pushing to higher redshifts we might hope to make better use of data at scales in the non-linear regime. Surveys such as SDSS already measure the power spectrum on  non-linear  scales, but are unable to make full use of this information. In principle, quasi-linear modes may be used to obtain cosmological information if they can be accurately modelled by, for example, higher order perturbation theory \cite{jeong2006,crocce2006}.  In Figure \ref{fig:kmax_ngal_plot}, we show the improvement in a survey that reaches to $z=3$ if modes out to $k_{\rm max}(\gamma)$ can be included.  The $n_s$ constraint is relatively insensitive to $k_{\rm max}$, while gains in $\alpha_s$ by a factor of a few can be achieved by accessing smaller scales.  This agrees with the findings of \cite{takada2006}.
As expected, the flattening of the contours with increasing $\gamma$ shows that to take advantage of the extra scales larger $n_{\rm gal}$ is required.

While an understanding of nonlinearities may improve our ability to use small scale modes, non-linear or non-local biasing may prevent this.  Non-linear biasing may be modeled via the halo occupation distribution (HOD) \cite{seljak2000} or via higher order perturbation theory \cite{mcdonald2006}. However,  either approach introduces additional free parameters that must be marginalized over, degrading parameter constraints.  Inclusion of higher order bias terms has been shown to have relatively little effect on baryon acoustic oscillations \cite{jeong2009}, where the oscillatory features are robust to changes in the broad band power, but may prove more important when constraining inflationary parameters.  At the level of the Fisher matrix, we have found that accounting for third order bias parameters, of the form detailed in \cite{mcdonald2006}, degrades the constraints on $n_s$ and $\alpha_s$ by a factor of $\sim2$.  Systematic effects from the non-linear biasing are beyond the scope of our analysis.  Additionally, non-local biasing can occur if, for example, the large scale features of reionization modulate the galaxy power spectrum via suppression of galaxies in ionized regions \cite{pritchard2007}.  This would also modify the broad band power of the galaxy power spectrum introducing systematic uncertainties.

\subsection{21~cm Experiments}
\label{sec:21cm}

In the preceding section, we have shown the need to utilise high-redshift galaxy surveys in order to push constraints on $\alpha_s$ to the $10^{-4}$ level.  These redshifts may also be probed with the redshifted 21 cm signal which, in the absence of ionization fluctuations, tracks the density field.  Several low-frequency radio interferometers including GMRT \cite{pen2009}, LOFAR\footnote{http://www.lofar.org/}, MWA\footnote{http://www.MWAtelescope.org/}, and PAPER \cite{parsons2009} are currently under construction and hope to make an initial detection of this signal. While these first generation experiments are unlikely to add much to our knowledge of cosmological parameters, future 21 cm experiments such as SKA\footnote{http://www.skatelescope.org/} have the potential to significantly improve upon our current knowledge of inflationary parameters.  The analysis of \cite{Mao:2008ug} showed that a futuristic 21 cm experiment, the Fast Fourier Transform Telescope (FFTT) \cite{tegmark2009}, could potentially achieve $\Delta\alpha_s=10^{-4}$, under optimistic assumptions about reionization. 

Such experiments can be focussed either at moderate redshifts $z=7-25$, where the effects of reionization and the first stars will serve as a major contaminant \cite{pritchard2008}, or at redshifts $z=25-50$, before star formation has begun.  Observing high redshifts requires removing foregrounds, especially galactic synchrotron emission which typically scales as $\nu^{-2.6}$ and dominates the signal.  We consider both regimes separately.  We also note that 21 cm intensity mapping \cite{wyithe2007,chang2008} potentially offers another way of going after the density power spectrum.  Since the challenges and redshifts associated with intensity mapping are largely the same as those associated with galaxy surveys, we will not consider them in detail.

\subsubsection{Cosmology from the Epoch of Reionization}

Observations of the 21 cm signal from the epoch of reionization (EoR) constrain a combination of the fluctuations in the density field and the ionization field.  Neglecting fluctuations in the 21 cm spin temperature (which is a reasonable, although not guaranteed to be safe \cite{pritchard2008}), the 21 cm brightness temperature power spectrum takes the form
\begin{equation}\label{ptb}
P_{T_b}=P_{\delta\delta}+2P_{\delta x}+P_{xx}-2\mu^2(P_{\delta \delta}+P_{x\delta})+\mu^4P_{\delta\delta}+P_{f(k,\mu)}.
\end{equation}
where $\mathbf{k}$ and the line of sight, $P_{\delta\delta}$, $P_{xx}$, and $P_{x\delta}$ are the power spectrum of the density field, the ionization field, and the density-ionization cross-correlation respectively, while $\mu$ is again the angle between the Fourier mode and the line of sight.  The final term $P_{f(k,\mu)}$ contains contributions higher than quadratic in the perturbations.  Although these would normally be neglected as small, during reionization fluctuations in the ionized fraction $x_i$ can be of order unity, so terms higher than quadratic contribute to the power spectrum.  These terms spoil the simple angular dependence expected from linear theory \cite{mcquinn2006} and potentially degrades our ability to separate astrophysics from cosmology beyond that considered in \cite{Mao:2008ug}.  Numerical simulations \cite{lidz2007,santos2008} show that higher order terms contribute significantly to the power spectrum on all scales once reionization is underway, and must therefore be included if cosmological parameters are to be correctly estimated.  

We model $P_{xx}$ and $P_{\delta x}$ using fitting functions and fiducial parameters from \cite{Mao:2008ug}.  These take the form
\begin{eqnarray}
P_{\delta x}&=& b_{\delta x}^2\exp[-\alpha_{x\delta}(k R_{x\delta})-(k R_{x\delta})^2]P_{\delta\delta}\nonumber\\
P_{xx}&=&b_{xx}^2[1+\alpha_{xx}(k R_{xx})+(k R_{xx})^2]^{-\frac{\gamma_{xx}}{2}}P_{\delta\delta},
\end{eqnarray}
where $b_{\delta x}$, $\alpha_{x\delta}$, $R_{x\delta}$, $b_{xx}$, $\alpha_{xx}$, $R_{xx}$ and $\gamma_{xx}$ are free parameters to be varied in our Fisher analysis.

In our Fisher analysis, we will consider two scenarios for the ionized fluctuations: an optimistic scenario where ionization contributions are neglected (OPT), and a case where only quadratic contributions to the power spectrum are included (MID).  We have separately considered the more detailed case in which higher order terms are considered.  These require considerably more work and appear to degrade the inflationary constraints by only an additional $\sim50\%$ over the MID case.  For this reason, we leave a more detailed discussion of the higher order terms to separate work.

\subsubsection{Fisher matrix formalism}

We perform a Fisher analysis of 21 cm experiments following \cite{bowman2006,mcquinn2006}.  The Fisher matrix for a 21 cm experiment is given by a sum over angular bins
\begin{equation}
F_{ij}=\sum_{\rm \mu} \frac{\epsilon k^3 V_{\rm survey}}{4\pi^2}\frac{1}{\sigma_P^2(k,\mu)}\frac{\partial P_{T_b}}{\partial \lambda_i}\frac{\partial P_{T_b}}{\partial \lambda_j},
\end{equation}
where the variance of a 21 cm power spectrum estimate for a single
$\mathbf{k}$-mode with line of sight component $k_{||}=\mu k$ is given by
\begin{equation}
\sigma_P^2(k,\mu)= \frac{1}{N_{\rm field}}\left[\bar{T}_b^2P_{21}(k,\mu)+T_{\rm sys}^2\frac{1}{B t_{\rm int}}\frac{D^2\Delta D}{n(k_\perp)}\left(\frac{\lambda^2}{A_e}\right)^2\right]^2 \, .
\end{equation}
This depends upon the system temperature $T_{\rm sys}$, the survey
bandwidth $B$, the total observing time $t_{\rm int}$, the conformal
distance $D(z)$ to the center of the survey at redshift $z$, the depth of
the survey $\Delta D$, the observed wavelength $\lambda$, and the effective
collecting area of each antennae tile $A_e$.  The effect of the
configuration of the antennae is encoded in the number density of baselines
$n(k_\perp)$ that observe a mode with transverse wavenumber $k_\perp$.  Observing a number of fields $N_{\rm field}$ further
reduces the variance.  We will consider only the case of a filled array and note that for baselines well below the cutoff due to the finite size of the array $n(k_\perp)\approx N_{\rm ant}\lambda^2/A_e=(A_{\rm tot}/A_{\rm e})(\lambda^2/A_e)$.  The response of a single dipole sets the minimum effective area $A_e=\lambda^2/4$.  These arrays are therefore usefully described in terms of $A_e$, $A_{\rm tot}$, $B t_{\rm int}$, and the redshift range covered.

\begin{figure}[tbp]
\begin{center}
\includegraphics[scale=0.85]{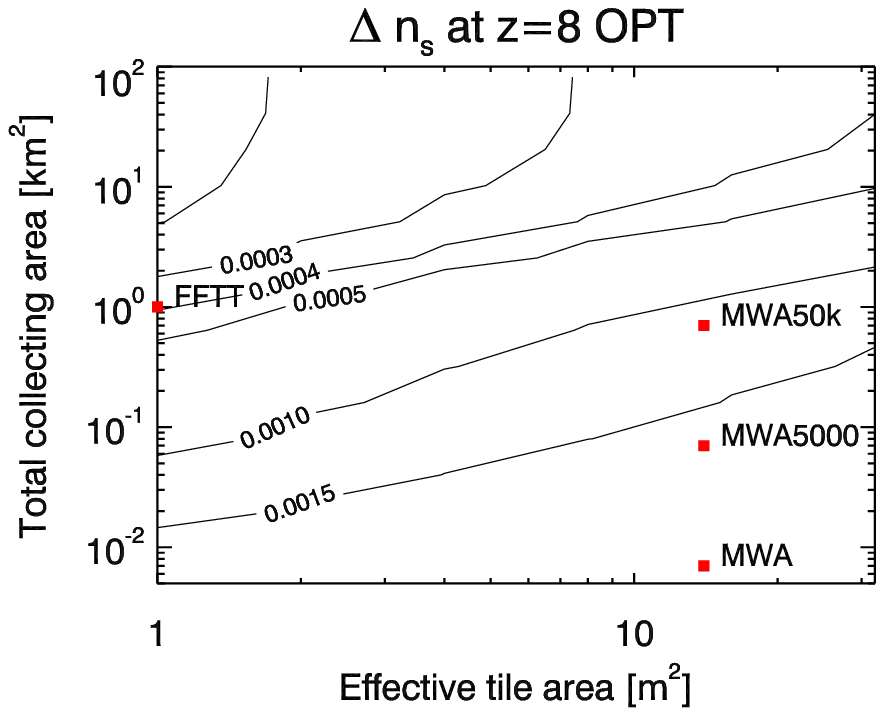}
\includegraphics[scale=0.85]{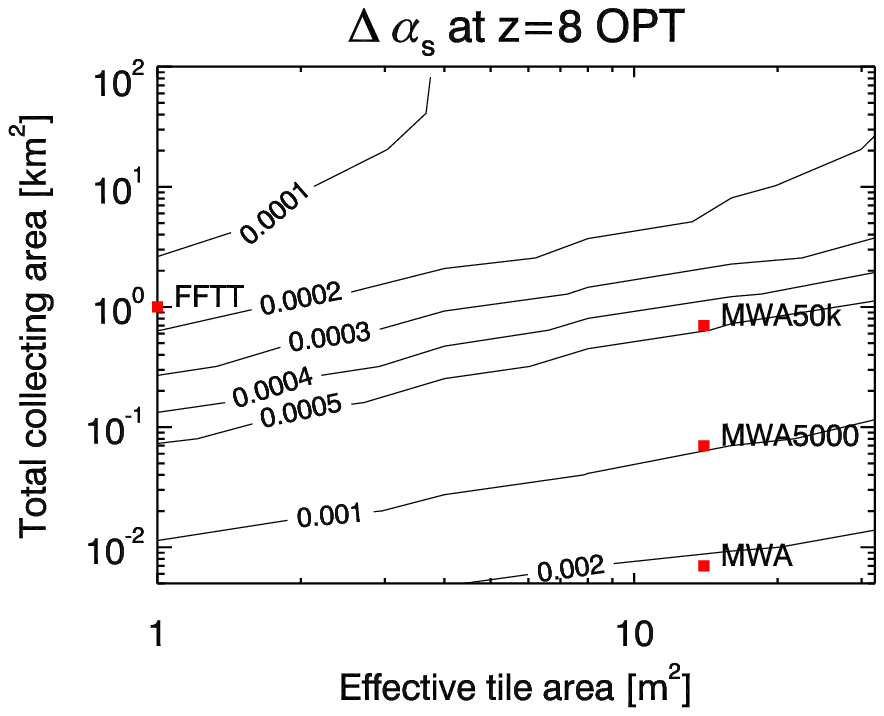}
\caption{Contour plot of $\Delta n_s$ (left panel) $\Delta\alpha_s$ (right panel) in the $A_{\rm e}-A_{\rm tot}$ plane at $z=8$ in the OPT scenario.}
\label{fig:aeff_atot_plot}
\end{center}
\end{figure}
\begin{figure}[tbp]
\begin{center}
\includegraphics[scale=0.85]{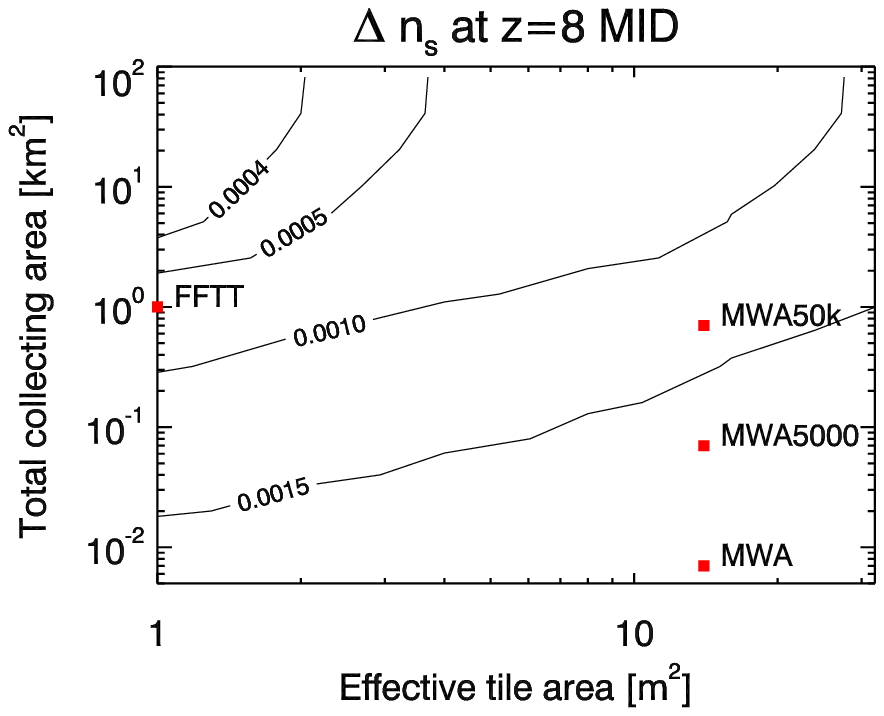}
\includegraphics[scale=0.85]{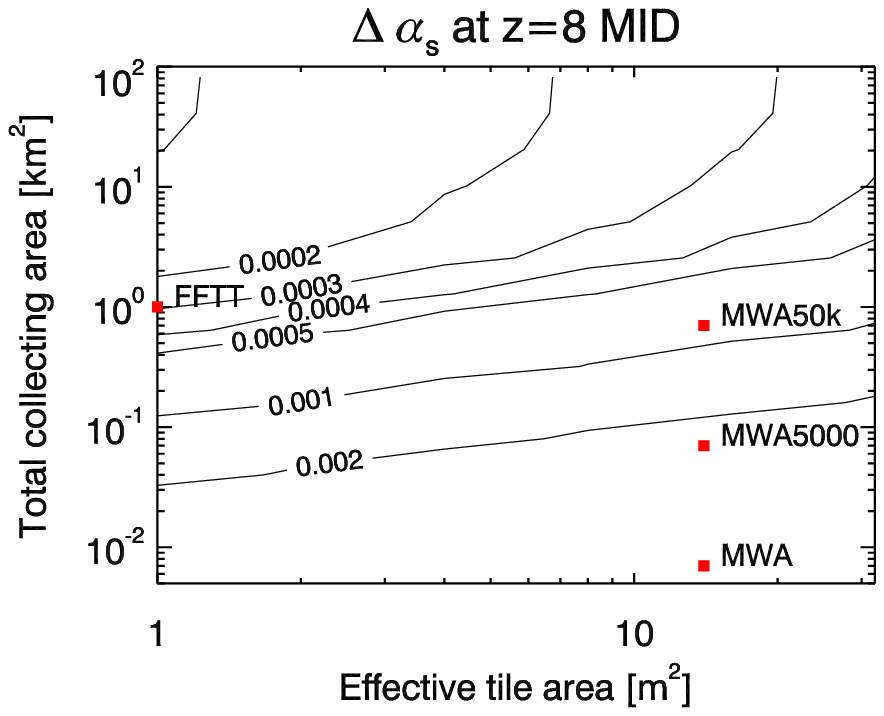}
\caption{Contour plot of $\Delta n_s$ (left panel) $\Delta\alpha_s$ (right panel) in the $A_{\rm e}-A_{\rm tot}$ plane at $z=8$ in the MID scenario.}
\label{fig:aeff_atot_plot_ion}
\end{center}
\end{figure}
The sensitivity of 21 cm experiments in the $A_{\rm e}-A_{\rm tot}$ plane for $B=8$ MHz, $N_{\rm field}=2$, and 4000 hours integration time is shown in Figure \ref{fig:aeff_atot_plot}.  $A_{\rm tot}$ affects both the sensitivity and the angular resolution, which controls the largest $k_\perp$ accessible.  $A_{\rm e}$ also impacts the sensitivity and controls the volume probed by the survey via the instantaneous field of view.  Once the experiment becomes sample variance limited the only way to improve sensitivity is via increasing the volume probed, i.e. by making $A_{\rm e}$ smaller.

It was shown in \cite{Mao:2008ug} that the two most important limiting factors for 21 cm observations are, respectively, modeling of the contribution to the power spectrum from ionization fluctuations and foregrounds.  In Figure \ref{fig:aeff_atot_plot}, we show contours in the case where ionization fluctuations are ignored (OPT), while in Figure \ref{fig:aeff_atot_plot_ion} we allow for quadratic terms (MID).  In the MID case constraints are degraded by a factor of $\sim2$ over the OPT case.  

In each of these figures, we mark the approximate position of several current and future 21 cm experiments.  We indicate MWA, built out of 500 antennae tiles with $A_{\rm eff}=14{\,\rm m^2}$ at $z=8$, and two possible successor instruments built  by increasing the number of tiles by a factor of 10 (MWA5000) and 100 (MWA50k).  We also mark the specifications of the proposed FFTT, composed of $10^6$ dipoles with $A_{\rm eff}=\lambda^2/4$.  We have assumed a compact array design for all of these, which is not actually how MWA is being built, although this give a small correction.  MWA50k and FFTT have similar collecting area, the boost from cross-correlating all of the dipoles makes FFTT much more sensitive to the power spectrum, both from a greater raw sensitivity and since it surveys a larger volume of sky.

As expected from our earlier discussion, the increased volume accessible from a full sky survey at $z=8$ allows constraints on the inflationary parameters at the level needed to detect $\alpha_s$ at levels which are characteristic of simple slow roll models, and  FFTT has the instrumental sensitivity required to actually make measurements at this level.  Allowing for ionization contributions degrades the sensitivity to the running, but the high sensitivity of the instrument means that even after this degradation,  interesting constraints on the running are still obtainable.  However, the need to model ionization contributions raises the possibility of systematic biases, especially at this high level of precision and it is still unknown whether future 21 cm experiments will be able to remove foregrounds and control systematics at the level needed to achieve this  sensitivity.

\subsubsection{21 cm cosmology from the dark ages}

We have seen that trying to obtain cosmology from the 21 cm signal during the epoch of reionization is complicated by the presence of fluctuations in the neutral fraction.  In principle, one can avoid this by going to redshifts $z\gtrsim30$, before star formation has begun.  Here brightness fluctuations are expected to trace the density as a result of collisional coupling \cite{lewis2007,pritchard2008}.  

There are two main challenges to accessing this redshift range.  The first is the Earth's ionosphere, which has a plasma frequency of $\sim10$ MHz leading to smearing of the signal and a resulting loss of angular resolution that prevents observations of the $z\gtrsim30$ regime from the ground \cite{jester2009}.  This requires that observations be conducted from space, including proposals based on the lunar surface.  The second challenge is that foregrounds  scale with frequency as $\nu^{-2.6}$ and become very large at the frequencies of interest, making observations significantly more difficult than during the epoch of reionization \cite{pritchard2008}.  Proposals for lunar arrays exist, such as LARC and DALI \cite{burns2009}, although these are geared towards arrays with size $\sim$km$^2$.  Since   low frequency dipole antennae can be little more than wires printed on a sheet of plastic a lunar array is less crazy that it might initially appear, and the proposed Ares V heavy launch vehicle could deliver $\sim 0.5$ km$^2$ of collecting area to the lunar surface.

\begin{figure}[tbp]
\begin{center}
\includegraphics[scale=0.85]{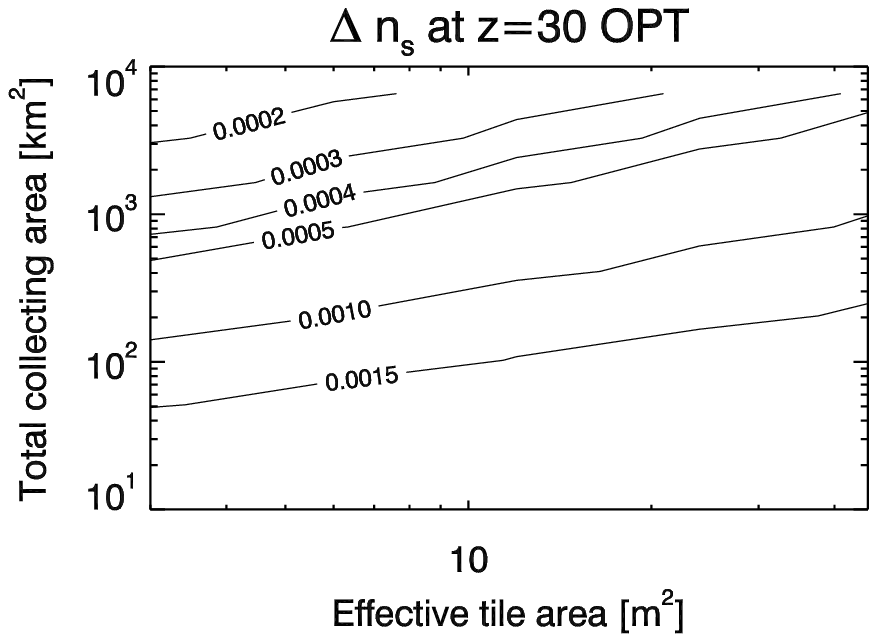}
\includegraphics[scale=0.85]{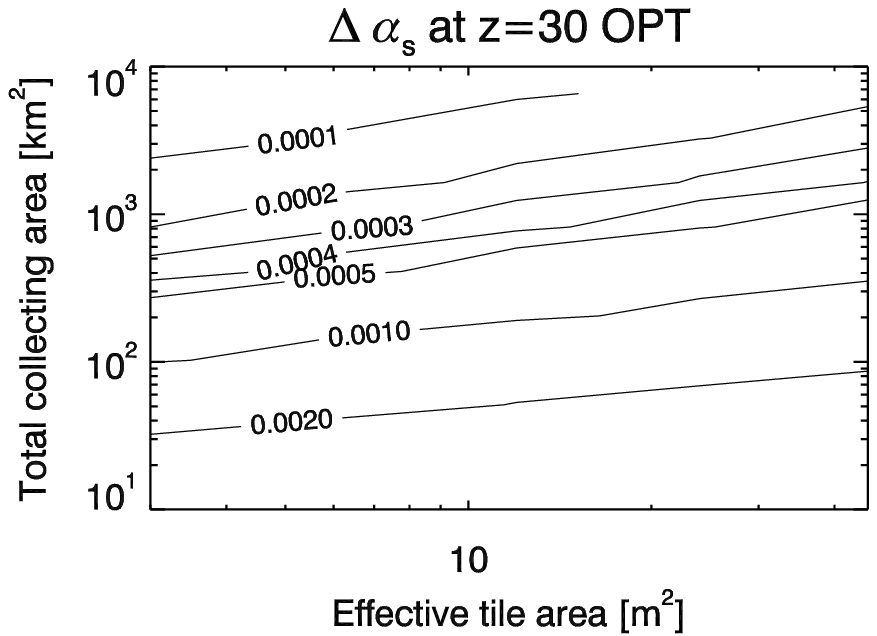}
\caption{Contour plot of $\Delta n_s$ (left panel) $\Delta\alpha_s$ (right panel) in the $A_{\rm e}-A_{\rm tot}$ plane at $z=30$ in combination with EPIC.}
\label{fig:aeff_atot_plot_z30}
\end{center}
\end{figure}
In Figure \ref{fig:aeff_atot_plot_z30}, we explore the requirements of measuring the tilt and running with a lunar array.  Given the time scale required for building such an array, we consider the inflationary constraints from a 21 cm instrument in combination with EPIC \cite{bock2009}, which by itself achieves $\Delta n_s=0.0018$ and $\Delta\alpha_s=0.0026$, in our calculations.  Arrays with collecting area of order 10 km$^2$ are required to measure the 21 cm signal and achieving a high signal of noise requires even larger collecting areas.  An FFTT like array with collecting area $10^3$ km$^2$ could measure the running at the level of $10^{-4}$ from the lunar surface, which is clearly an extremely futuristic proposal. Smaller arrays might still be useful for constraining inflation via observations of the large scale power spectrum, which can be used to constrain  compensated isocurvature modes \cite{gordon2009}.

We list  the constraints on the inflationary sector that can be obtained with  our fiducial versions of CIP and FFTT in Table \ref{tab:constraints_21cm}.  CIP3 (CIP4) assumes a galaxy survey over 1000 (10000) sq. deg. with redshift bins at $z=3$, 4, and 5 achieving galaxy densities of $n_{\rm gal}=8\times10^{-3}$, $4\times10^{-3}$, and $1\times10^{-3}$$\,{h^3{\rm Mpc^3}}$ respectively.  For FFTT, we assume $10^6$ dipoles with $A_{\rm eff}=\lambda^2/4,$ for a collecting area of $A_{\rm tot}\approx$1 km$^2$ at $z=8$.  For this illustration, we take the fiducial inflationary parameters corresponding to natural inflation with $f=\sqrt{8\pi}$ and $N=51$.

\begin{table}[tbp]
\caption{Inflationary parameter constraints for CIP and FFTT. The first block gives forecasts for CIP and FFTT on their own, the second block gives forecasts for Planck, Planck+CIP and Planck + FFTT.  The Fisher forecasts are performed with the fiducial parameters  $n_s =0.95$, $\alpha_s = -0.0005$, and $r=.0045$.  }
\begin{center}
\begin{tabular}{ccccc}
\multicolumn{2}{c}{Experiment} & $\Delta n_s$ & $\Delta \alpha_s$ & $\Delta  r$\\
\hline
\hline
   \multicolumn{2}{c}{CIP3} & 0.0089 & 0.0028 & - \\
 \multicolumn{2}{c}{CIP4} & 0.0028 & 0.0009 & - \\
\multirow{2}{*}{FFTT}&OPT & 0.0011 & 0.00023 & - \\
&MID & 0.00082 & 0.00032 & - \\
 \hline \hline
\multicolumn{2}{c}{Planck} & 0.0032 & 0.005 & 0.058 \\
\hline
\multicolumn{2}{c}{+CIP3} & 0.0019 & 0.0011 & 0.05 \\
\multicolumn{2}{c}{+CIP4} & 0.0011 & 0.0006 & 0.048 \\
 \hline
\multirow{2}{*}{+FFTT}&OPT & 0.00034 & 0.000095 & 0.048 \\
&MID & 0.00067 & 0.00028 & 0.048 \\
\end{tabular}
\end{center}
\label{tab:constraints_21cm}
\end{table}%
 
In principle, these experiments are capable of precise measurements of the tilt and borderline detections of the running expected for simple slow roll models. However, once systematic effects from biasing or reionization are taken into account it is apparent that a precision measurement of the running is an extremely challenging project.    The best hope for improving this picture is probably to correctly model modes within the non-linear regime, allowing them to included in the analysis.  Nonetheless,  it is clear that there is a sufficient number of comoving modes within the visible Universe that a detection of the inflationary running is feasible.

\begin{figure}[tbp]
\begin{center}
\includegraphics[ width = 15cm]{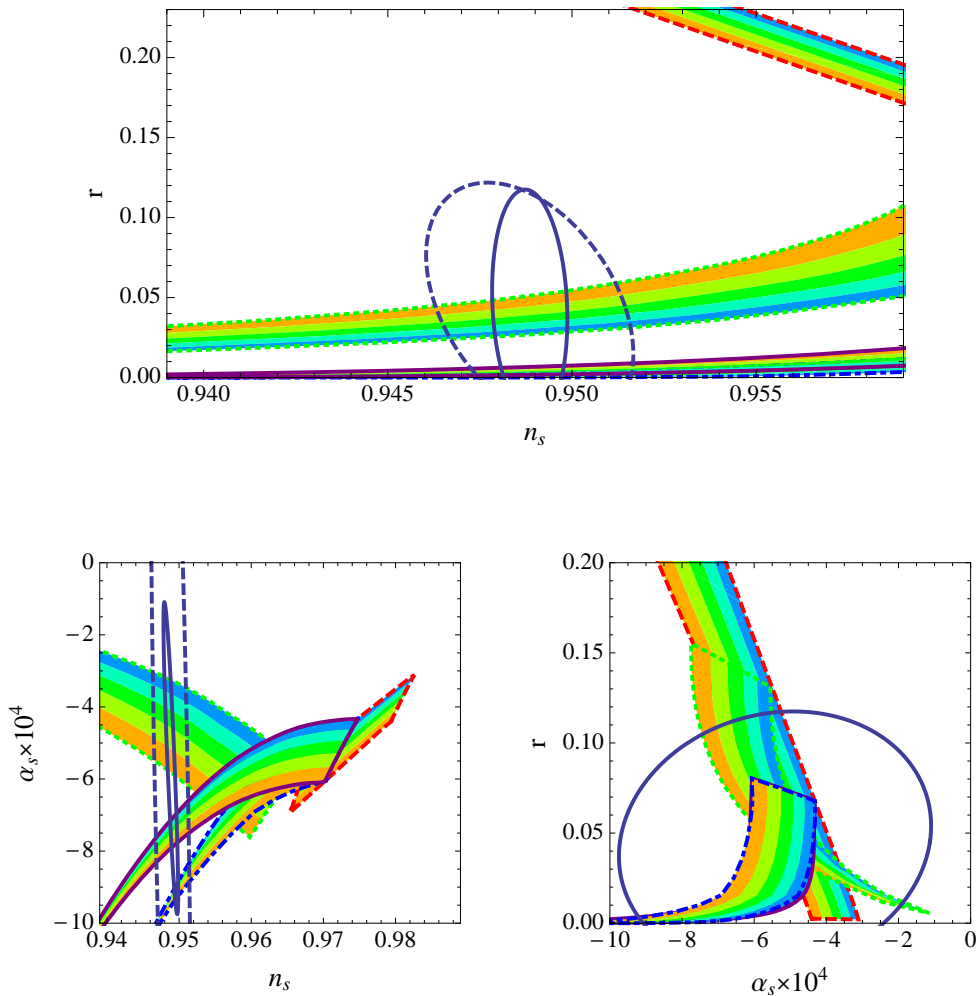} 
\caption{68-\% confidence ellipses for Planck+CIP (dashed ellipse) and Planck+FFTT (solid ellipse)  drawn on the  inflationary parameter space.  The Fisher analysis is performed assuming natural inflation, and the $(n_s,\alpha_s)$ plot only includes models with $r<0.12$. We see that in this scenario, all $\phi^n$ models are eliminated at high significance, but that we cannot use constraints on $\alpha_s$ to break the degeneracy between the remaining models.  \label{fig:natural_degeneracy}}
\end{center}
\end{figure}

\section{Constraining the post inflationary universe}
\label{sec:reheatconstrain}

The previous section  assessed the ability of galaxy surveys and 21 cm experiments  to detect the running predicted by canonical single field inflationary models.  If the primordial perturbations were generated by one of these models, the analysis of the previous section shows that a direct measurement of $\alpha_s$ is possible with future observations, given exquisite control of foregrounds and systematics.   Figure \ref{fig:natural_degeneracy} shows the error forecasts for measurements of $n_s$, $r$ and $\alpha_s$, with  a ``modest'' (Planck-level) constraint on the tensor signal, and tight constraints on $n_s$ and $\alpha_s$.  For this scenario, $\phi^n$ models are excluded at high significance. However, while FFTT is forecast to provide a marginal detection of $\alpha_s$, it cannot break the degeneracy between the remaining hilltop models and natural inflation.  Conversely, a post-Planck polarization mission would detect the tensor signal from natural inflation, and exclude the hilltop models.

\begin{table}[tbp]
\caption{Constraints on reheating temperature and inflationary model parameters from Planck,  CIP3 and FFTT (MID) for both natural and $\phi^n$ inflation.  Rows with ``-" give error forecasts for the complementary parameter, with the ``-'' parameter assumed to be specified in the prior. In each case the final row gives a constraint assuming that $r$ is measured to $\pm 0.01$. }
\begin{center}
\begin{tabular}{|c|cc|cc|}
\hline
&\multicolumn{2}{c|}{Natural} &\multicolumn{2}{c|}{$\phi^n$}\\
& $N$ & $f$ & $N$ & $n$\\
\hline
fiducial values & 51 & $\sqrt{8\pi}$ & 51 & 2\\
\hline
Planck & 5.1 & - & 3.6& -\\
 & - & 0.33  & - & 0.25\\
 & 14.5 & 0.93  & 19.7 & 1.4\\
 + $\sigma_r=0.01$ &3.5 & 0.26 &8.6 & 0.41\\
\hline
CIP+Planck & 1.69 & - & 1.2 & -\\
 & - & 0.11  & - & 0.09\\
 & 13.7 & 0.87  & 14.5 & 1.14\\
 + $\sigma_r=0.01$ &2.8 & 0.18 &3.96 & 0.27\\
\hline
FFTT+Planck & 0.41 & - & 0.29 & -\\
 & - & 0.027  & - & 0.024\\
 & 7.0 & 0.45  & 11.0 & 0.91\\
 + $\sigma_r=0.01$ &2.5 & 0.17 &2.95 & 0.24\\
 \hline
\end{tabular}
\end{center}
\label{tab:model_constraints}
\end{table}%

However, as discussed in Sections 2 and 3, the running makes its presence felt at much lower levels of precision by inducing a correlation between $n_s$ and the post-inflationary expansion history.  Consequently, {\em given an explicit inflationary prior\/}, accurate constraints on the power spectrum via large scale structure surveys provide tight constraints on $N$ and the four variables needed to describe the primordial perturbations are typically functions of a smaller number of free parameters in the inflationary potential. The reduction in the dimensionality of the parameter space imposed by an inflationary prior thus allows us to constrain $N$, even when we cannot measure $\alpha_s$ with confidence.   Written in terms of the inflationary parameters, the Fisher matrix becomes
\begin{equation}
F^{\rm new}_{nm}=\sum_{ij}\frac{\ud q_i}{\ud p_n}F_{ij}\frac{\ud q_j}{\ud p_m},
\end{equation}
where the $q_i$ are the spectral parameters $n_s$, $\alpha_s$, and $r$ while the $p_n$ are the free  parameters in the specific inflationary model, including the number of e-foldings $N$. Table \ref{tab:model_constraints} lists the constraints on inflationary model parameters and $N$ that can be expected from Planck,  CIP and FFTT (in the MID scenario).  In practice, these forecasts are approximate since the inflationary prior puts sharp cuts on the parameter space, while the Fisher matrix assumes a Gaussian likelihood.

\begin{figure}[tbp]
\begin{center}
\includegraphics[scale=1]{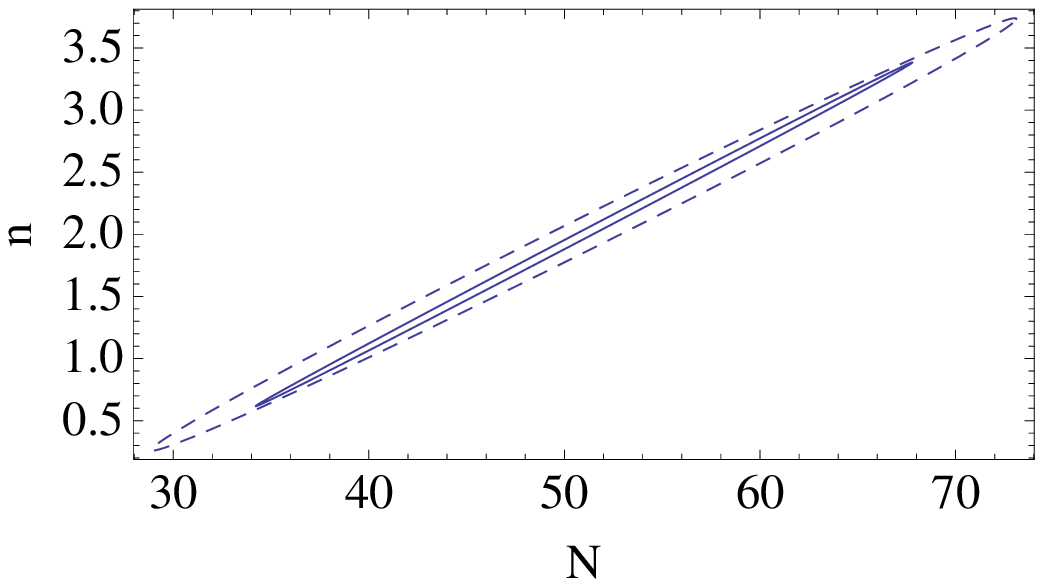} \\  \mbox{} \\
\includegraphics[scale=1]{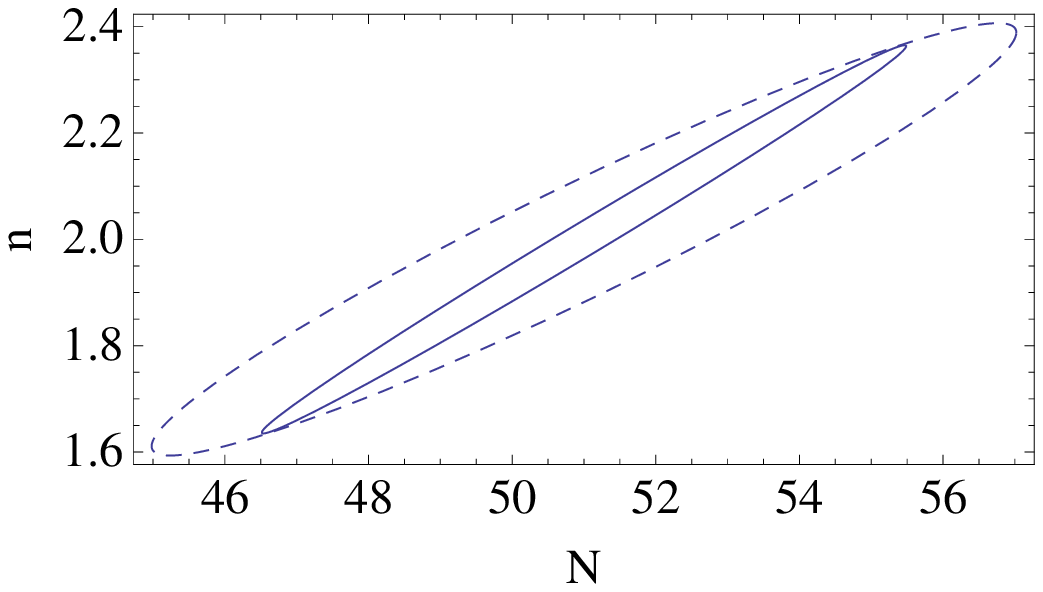}
\caption{ Top: 68-\% confidence ellipses for Planck+CIP (dashed ellipse) and Planck+FFTT (solid ellipse) for $\phi^n$ inflation mapped into the $N-p$ plane.  Bottom: As above, but assuming $r$ is measured to within 0.01.}
\label{fig:XPellipse}
\end{center}
\end{figure}

Given a tight prior on the inflationary potential, we see that  $N$ is likely to be well constrained by Planck, so that Planck could begin to distinguish between quadratic inflation followed quickly by an unbroken period of radiation domination, and a scenario that involved thermal inflation.  This knowledge is of significant value to particle theorists and inflationary model builders, since it implies a non-trivial correlation between the assumed form of the inflationary potential and the properties of the post-inflationary universe.  Beyond Planck, FFTT and CIP would provide exquisite constraints on the post-inflationary universe, for a given inflationary scenario.   

Clearly, if the inflaton potential has more free parameters, constraints on $N$ weaken substantially.  For the examples here, estimating  both $f$ or $n$ (for natural and $\phi^n$ inflation, respectively)  and $N$ from data  leads to substantial play in the allowed values. However, a strong constraint on $r$  and a  CIP-class measurement of large scale structure is sufficient to put a tight constraint on both the shape of the potential and $N$. Figure \ref{fig:XPellipse} shows the constraint forecasts for $n$ and $N$, for $\phi^n$ inflation.  There is considerable degeneracy between these two parameters that neither CIP or FFTT are able to break on their own.  However, a precision  measurement of $r$ breaks this degeneracy, so that the parameters are accurately determined by the data.

\section{Discussion}
\label{sec:discussion}

We have surveyed the impact of a running  spectral index on the predictions of inflationary models. Beyond the obvious role of $\alpha_s$ as a free parameter in its own right, a running index and unknown expansion history of the post-inflationary universe induce a substantial theoretical uncertainty in the predictions of inflationary models.  Physically, the field value at which perturbations are generated is sensitive to the post inflationary expansion history, which determines that rate at which modes reenter the horizon after inflation.   Thus if the universe does not immediately thermalize at the end of inflation, the pivot scale is effectively shifted (relative to that defined for instantaneous reheating) to  $k'_{\star} = k_{\star}e^{\Delta N}$.     

If $n_s$ was actually constant (as is the case for power-law inflation) changing $k_{\star}$ would not change $n_s$. However, even simple inflationary models have $\alpha_s$  large enough  to ensure that the post-inflationary expansion history has a nontrivial impact on the measured value of the spectral index.     Making the conservative assumption that inflation is followed by an effective matter dominated phase with thermalization occurring at the TeV scale or above, the resulting uncertainty in $n_s$ is comparable to the statistical uncertainty expected from Planck.  Conversely, with only minimal theoretical assumptions about the properties of the post-inflationary universe the corresponding uncertainty in $n_s$ is similar to the statistical uncertainty found in WMAP constraints.  

This ambiguity arises because inflationary models are typically specified solely in terms of their potential, rather than being embedded in a larger theory of particle physics. Consequently, positing an inflationary model does not specify the properties of particle physics from TeV to GUT scales, or the couplings between the inflaton and the ``rest of particle physics''.  With this information, the post-inflationary expansion history of the universe could be calculated, and the uncertainty in the inflationary spectrum removed.  

Our analysis shows that futuristic large scale structure or 21cm measurements  will detect $\alpha_s$ at the level predicted by simple inflationary models, confirming previous work in this area. However, a measurement that is accurate enough to distinguish {\em between\/} the values of $\alpha_s$ predicted by these models will be extremely challenging.  The ultimate limits to these observations will likely be set by systematic effects, so that we have dwelt only upon best possible statistical limits, leaving the issue of systematics aside.

Conversely, the uncertainty in $n_s$ induced by $\alpha_s$ will be important to all analyses of the inflationary parameter space following Planck, even if $\alpha_s$ is not directly detected.  This apparent paradox is resolved when we recall that measuring $\alpha_s$ directly requires estimating at least four independent spectral parameters (in addition to other concordance variables), namely $A_s$, $r$, and $n_s$, in addition to  $\alpha_s$.  Conversely, explicit inflationary models typically have fewer free parameters,  effectively correlating one or more of the otherwise independent spectral parameters.   Consequently, given a {\em specific inflationary potential\/}, we can constrain the integrated expansion history of the post-inflationary universe, as described in Section 5.   

Intriguingly,   many supersymmetric scenarios predict that the primordial universe undergoes a long matter dominated phase, due to the presence of ``light'' moduli whose energy density scales like non-relativistic matter.  In some scenarios these moduli are erased by thermal inflation \cite{Lyth:1995ka}; in others they decay of their own accord in such a way as to avoid disrupting nucleosynthesis \cite{Acharya:2010af}.   In both cases, the number of e-folds required to match the pivot scale to the moment it left the horizon during inflation is substantially different from that of a universe which is thermalized  throughout the post-inflationary epoch. Consequently, in the near future astrophysical observations will determine which inflationary scenarios are compatible with these common supersymmetric scenarios, and which are not. 

The Large Hadron Collider promises to extend our understanding of particle physics to the TeV scale, while direct detection experiments are putting significant constraints on many dark matter  models.  If these experiments reveal the mechanisms responsible for setting the present-day baryon and dark matter fractions, we will gain significant new windows into the very early universe.  Given the huge range of energies that can lie between the inflationary and TeV scales, along with the lack of direct probes of particle physics at these energies, the post-inflationary era amounts to a fundamental ``dark age'', where both the cosmological evolution and fundamental laws of physics are unknown.   We have seen here that the relationship between $n_s$  and the post-inflationary expansion history induced by  $\alpha_s$ may help to illuminate this currently mysterious epoch.  

\ack    PA and RE are supported in part by the United States Department of Energy, grant
DE-FG02-92ER-40704. RE is supported by an NSF Career Award PHY-0747868.  JRP is supported by  NASA through Hubble Fellowship grant HST-HF-51234.01-A awarded by the Space Telescope Science Institute, which is operated by the Association of Universities for Research in Astronomy, Inc., for NASA,
under contract NAS 5-26555.  This work was supported in part by NSF grant AST-0907890 and NASA grants NNX08AL43G and NNA09DB30A for AL.  RE and JRP  thank the Aspen Center for Physics for hospitality during  part of this work.   We thank Will Kinney and Hiranya Peiris  for a number of useful conversations.

\section*{References}
\bibliographystyle{h-physrev3}
\bibliography{fisher,obs}{}


\end{document}